\documentclass[traditabstract]{aa}
\usepackage{graphicx}
\usepackage{wasysym}
\usepackage{txfonts}
\usepackage{natbib}
\usepackage{ulem}
\bibpunct{(}{)}{;}{a}{}{,} 

\def\cone {\ifmmode{{\rm C}{\rm \small I}(^3\!P_1\!-^3\!P_0)}
     \else{C\ts {\scriptsize I}{\small$(^3\!P_1\!-^3\!\!\!P_0)$}}\fi}
\def\ctwo {\ifmmode{{\rm C}{\rm \small I}(^3\!P_2\!-^3\!P_1)}
     \else{C\ts {\scriptsize I}{\small$(^3\!P_2\!-^3\!\!\!P_1)$}}\fi}

\def\tex {\ifmmode{{T}_{\rm ex}}\else{$T_{\rm ex}$}\fi}
\def\tmb {\ifmmode{{T}_{\rm mb}}\else{$T_{\rm mb}$}\fi}
\def\ci     {\ifmmode{{\rm C}{\rm \small I}}\else{C\ts {\scriptsize I}}\fi}
\def\hi     {\ifmmode{{\rm H}{\rm \small I}}\else{H\ts {\scriptsize I}}\fi}
\def\hh     {\ifmmode{{\rm H}_2}\else{H$_2$}\fi}

\def\ts     {\thinspace}
\def\kms    {\ifmmode{{\rm \ts km\ts s}^{-1}}\else{\ts km\ts s$^{-1}$}\fi}
\def\msol   {\ifmmode{{\rm M}_{\odot}}\else{M$_{\odot}$}\fi}
\def\lsol   {\ifmmode{{\rm L}_{\odot}}\else{L$_{\odot}$}\fi}
\def\zsol   {\ifmmode{{\rm Z}_{\odot}}\else{Z$_{\odot}$}\fi}

\begin{document}

\title{A DEBRIS Disk Around The Planet Hosting M-star GJ~581 Spatially  Resolved with Herschel
\thanks{{\it Herschel} in an ESA space observatory with science
instruments provided by European-led Principal Investigator
consortia and with important participation by NASA.}}

\subtitle{}

\author{
J.-F. Lestrade \inst{1}
\and
B. C. Matthews \inst{2,8}
\and
B. Sibthorpe \inst{3}
\and
G.M. Kennedy \inst{4}
\and
M. C. Wyatt  \inst{4}
\and
G. Bryden \inst{5}
\and 
J. S. Greaves \inst{6} 
\and 
E.~Thilliez \inst{1}
\and
Amaya Moro-Mart\'in  \inst{7}
\and 
M. Booth \inst{8}
\and 
W.R.F. Dent  \inst{9}
\and 
 G. Duch\^ene  \inst{10, 11}
\and 
 P. M. Harvey  \inst{13}
\and 
J.~Horner \inst{14}
\and
P.~Kalas   \inst{10,12}
\and
JJ~Kavelaars \inst{2,8}
\and  
N.M.~Phillips  \inst{9}
\and 
D.R. Rodriguez\inst{15}
\and
K. Y. L. Su \inst{16}
\and 
D. J. Wilner   \inst{17}
}
\institute{Observatoire de Paris, CNRS, 61 Av. de l'Observatoire,  F-75014, Paris, France
\and
Herzberg Institute of Astrophysics (HIA), National Research Council of Canada, Victoria, BC, Canada               
\and
UK Astronomy Technology Centre (UKATC), Royal Observatory Edinburgh, Blackford Hill, Edinburgh, EH9 3HJ, UK
\and
Institute of Astronomy (IoA), University of Cambridge, Madingley Road, Cambridge, CB3 0HA, UK
\and
Jet Propulsion Laboratory, California Institute of Technology, Pasadena, CA 91109, USA
\and
School of Physics and Astronomy, University of St. Andrews, North Haugh, St. Andrews, Fife, KY16 9SS, UK
\and
Department of Astrophysics, Center for Astrobiology, Ctra. de Ajalvir, km 4, Torrej´on de Ardoz, 28850, Madrid, Spain
\and
Dept. of Physics \& Astronomy, University of Victoria, Elliott Building, 
3800 Finnerty Rd, Victoria, BC, V8P 5C2 Canada
\and
ALMA JAO, Av. El Golf 40 - Piso 18, Las Condes, Santiago, Chile
\and
Astronomy Department, UC Berkeley, 601 Campbell Hall, Berkeley CA 94720-3411, USA 
\and
UJF-Grenoble 1 / CNRS-INSU, Institut de Plan\'etologie et d'Astrophysique de Grenoble (IPAG) UMR 5274, Grenoble, F-38041, France
\and
SETI Institute, 515 North Whisman Road, Mountain View, CA 94043, USA
\and
Astronomy Department, University of Texas at Austin, 1 University Station C1400, Austin, TX 
\and
Department of Astrophysics and Optics, School of Physics, University of New South Wales, Sydney 2052, Australia
\and
Departamento de Astronomía, Universidad de Chile, Casilla 36-D, Santiago, Chile 
\and
Steward Observatory, University of Arizona, 933 North Cherry Avenue, Tucson, AZ 85721, USA
\and
Harvard-Smithsonian Center for Astrophysics, 60 Garden Street, Cambridge, MA 02138, USA
}

\offprints{J-F. Lestrade, \email{jean-francois.lestrade@obspm.fr}}

\date{Received 3 September 2012 ; accepted 5 November 2012}

\titlerunning{Resolved Disk Around GJ~581}

\abstract{Debris disks have been found primarily around intermediate and solar mass stars (spectral
types A-K) but rarely around low mass M-type stars. We have  
spatially resolved a debris disk around the remarkable M3-type star GJ~581 hosting multiple planets using  
deep PACS images at 70, 100 and 160~$\mu$m as part of the DEBRIS Program on the {\it Herschel Space Observatory}.
This is the second spatially resolved debris disk found around an M-type star, after the one surrounding 
the young star AU~Mic (12~Myr). However,
GJ~581 is much older (2-8~Gyr), and is X-ray quiet in the ROSAT data.   
We fit an axisymmetric model of the disk to the three PACS images and found that the best fit model is 
for  a  disk extending radially from $25\pm12$~AU to more than 60~AU.
Such a cold disk is reminiscent of the Kuiper Belt but it surrounds a low mass star ($0.3~M_{\odot}$) 
and its  fractional dust luminosity  $L_{dust}/L_*$ of  $\sim 10^{-4}$ is much higher.
The inclination limits of the disk found in our analysis make the masses of the planets
small enough to ensure the long-term stability of the system according to some dynamical simulations.
The disk is collisionally dominated down to submicron-sized grains and 
the dust cannot be expelled from the system by radiation or wind pressures because of the low luminosity 
and low X-ray luminosity  of  GJ~581.
We suggest that the correlation between low-mass planets and  debris disks recently found 
for G-type stars also applies to M-type stars.
Finally, the known planets, of low masses and orbiting within 0.3~AU from the star, cannot dynamically perturb 
the disk over the age of the star, suggesting that an additional 
planet exists at larger distance that is stirring the disk to replenish the dust. 
}


\maketitle


 
\section{Introduction} \label{intro}

A debris disk around a main sequence star is a collection of small 
bodies left over from the planet formation process. 
In our Solar system, the Asteroid belt 
and Edgeworth-Kuiper Belt are the two best known reservoirs of objects 
that remain from the planet formation process and range  in size 
from hundreds of kilometers in diameter to meter-scale bodies \citep[e.g.][]{Jewi00, Shep06}. 
Such reservoirs  are highly sculpted by the evolution of the planetary system 
in which they form \citep[e.g.][]{Peti01, Morb05, Lyka09}, 
and contain objects whose accretion was stymied by the formation and migration of giant planets 
in the system, or simply occurred too slowly for them to grow larger. 
Since debris disks contain a vast number of objects on very similar orbits, 
they experience a continual collisional grinding which produces and continually replenishes 
a population of dust. This dust allows us to directly detect debris disks around other stars 
in two ways. The dust is heated by radiation from the central star, and therefore 
emits thermal radiation with a temperature characteristic of its 
distance from its host star  \citep[e.g.][]{Auma84, Grea05}. In addition, the 
smallest grains of dust can efficiently scatter the light of the host 
star \citep[e.g.][]{Smit84, Kala05}.
The physical and observational  properties of debris disks  were defined by \citet{Lagr00}, and 
their studies were recently reviewed by \citet{Wyat08} and \citet{Kriv10} and will  eventually place  
our Solar system in context  \citep{Grea10}.

Almost all debris disks detected  by the  satellites {\it IRAS, ISO} and {\it Spitzer} \citep{Bryd06, Su06, Tril08}, 
HST \citep{Goli11}, and  ground-based telescopes \citep{Wyat08}  surround A-type and F,G,K-type stars  
despite several deep surveys of large samples of M~stars conducted from mid-IR 
to submillimeter  wavelengths \citep{Plav05, Lest06, Gaut07, Lest09, Aven12}. 
Currently, among the nearby M-stars, the only spatially resolved debris disk is around 
the very young M1~star AU~Mic \citep{kala04, Liu04, Kris05, Wiln12} which has been modeled by \citet{Auge06} and \citet{Stru06}.  
In addition, there are a few candidate disks with excesses above photospheric level
\citep[e.g.][]{SmithPS06, Lest06, Plav09}.
Finally, in the cluster NGC2547 ($\sim$40~Myr old and $\sim$ 433~pc), deep {\it Spitzer MIPS} observations have 
revealed 11 M-stars with 24~$\mu$m excesses above photospheric level and no excess at  70~$\mu$m; these observations have been interpreted
 as warm dust in debris disks \citep{Forb08}.

The fact that debris disks are more seldomly observed among M-stars than around 
higher-mass stars seems surprising at first, since all spectral types have similar detection 
rates of protoplanetary disks in the earlier stage of their evolution, according to observations of low density clusters like Taurus-Auriga and
$\rho$~Oph  \citep[e.g.,][]{Andr05}. However, in high density clusters like Orion, 
external photoevaporation by intense FUV radiation field can severely limit the production of planetesimals 
around low mass M-stars on a timescale shorter than $\sim$10~Myr \citep{Adam04}. 
Another hazard for M-stars during the first $\sim$100~Myr is close stellar flybys, when co-eval 
stars are still in the expanding cluster of their birth and strongly interacting with each other.
During these early close stellar flybys, planetesimals are stripped from disks, and this is more severe for 
disks around low mass stars in high stellar density  clusters like Orion according to 
simulations \citep{Lest11}.
 
Recently,  \citet{Wyat12}  have found evidence of 
the prevalence of debris disks in low-mass planetary systems (also Moro-Mart\'in et al. in prep) 
and suggest that this correlation could arise because such planetary systems 
are dynamically stable over Gyr timescales.        
Recent observations show that low-mass planets are more  abundant  
among M-stars than around the other stars \citep{Bonf11, Howa11}. 
Hence, if the correlation between debris disks and low-mass planets for G-stars 
applies to M-stars, then debris disks should be relatively common around them, 
in contrast to a paucity of detections.

However, debris disks around M-stars are harder to detect than around more massive stars 
at the same distance simply because they are less luminous, meaning that the dust within experiences 
significantly less heating. Therefore, to detect the same disk around a later type star 
requires deeper observations. M-star debris disks may also be less
detectable  because additional grain removal processes are operating. For example, a physical
pecularity of  M-stars is that they are structurally different from solar-type stars. Their interiors 
have deep convective zones $-$ fully convective for M3 spectral type and later $-$ that
produce strong coronal magnetic fields  responsible for their optical/radio flares and  X-ray emission  
\citep{Hawl00}.
This activity  generates  also stellar winds of energetic  particles  \citep{Warg01} which 
might dominate the circumstellar grain removal processes for a large fraction of the star lifetime 
\citep{Plav05}. 

This paper describes observations carried out as part of the Key Program DEBRIS 
(Disc Emission via a Bias-free Reconnaissance
in the Infrared/Sub-mm)  on the {\it Herschel Space
Observatory} \citep{Pilb10}. DEBRIS is an unbiased flux-limited survey  to search 
for dust emission at $\lambda=100$ and 160$\mu$m  toward
the nearest $\sim$89 stars of each spectral type A,F,G,K,M as 
evidence of debris disks (see \citet{Matt10} and \citet{Phil10} for the sample description). 
 For selected targets, complementary {\it Herschel} observations at 70, 250, 350, 500~${\mu}$m 
were also conducted. The first results of this program have already shown that these observations 
can detect disks down to much fainter levels than  previously achieved, and moreover can spatially 
resolve debris disks at far-IR wavelengths \citep{Matt10, Chur11, Kenn12, Wyat12, Kenn12b, Boot12, Broe12}. 

 As part of  this survey, we have  spatially resolved 
 a  disk around  the  M3 spectral type star GJ~581 at $\lambda$=70,  100, and  160~$\mu$m. 
 Hence, this is the second resolved 
 debris disk around an M-star, but, in contrast to the star AU~Mic which 
 is young \citep[12~Myr, ][]{Zuck04}, 
 GJ~581 is old (2-8~Gyr, see \S~\ref{par}). Also, GJ~581 is surrounded 
 by at least four low mass planets with minimum masses of 1.9, 15.6, 5.4, and 7.1~M$_{\oplus}$, 
 orbital radii of 0.03, 0.04, 0.07, and 0.22~AU,  and eccentricities between 0.0 and 0.32, 
 detected by radial velocity measurements \citep{Bonf05, Udry07, Mayo09, Forv11}. 
 All these planets are within  the tidal lock region of this M3 spectral type star ($<$0.25~AU) and hence
 are expected to be synchronously rotating and potentially undergoing atmospheric 
 instabilities \citep{Word11, kite11}. 
 Planets GJ~581c and d are near and in the conventionally defined Habitable Zone  \citep{Sels07}, respectively.
The presence of one or two  additional planets in the system is debated \citep{Vogt10, Forv11, Vogt12}.

  In this paper, we describe the {\it Herschel} observations of  GJ~581   
  as well as archival MIPS and IRS data from {\it Spitzer}, and NICMOS data from {\it HST} 
  in \S~\ref{obs}.  The stellar parameters of GJ~581 used are  in \S~\ref{par}.
  Reconnaissance of a cold debris disk around G~J581 in the three PACS images at 70, 100, and 160~$\mu$m
  and in the presence of background sources contaminating the field is described  in \S~\ref{images}. 
  Modeling of these images  to determine 
  the spatial distribution of the emitting dust  is described  in \S~\ref{mod}.
  The spectral energy distribution SED including the IRS spectrum of GJ~581 and 
  modeling of a hypothetical second component of warm dust  are described in  \S~\ref{sed}.
  An upper limit on the brightness of scattered light using the NICMOS image is discussed in  \S~\ref{Nicmos}. 
  Physical conditions in the disk and its relationship with the planetary system around  GJ~581 are discussed
  in \S~\ref{discussion}.

\section{Observations}  \label{obs}

\subsection{\bf Herschel }

GJ~581 was initially observed with PACS (Photodetector and Array Camera \&
Spectrometer, \citet{Pogl10}) on 11 August 2010 using the standard DEBRIS observing strategy, 
and  a resolved disk was tentatively detected at 100 and 160~$\mu$m.  
We then acquired deeper PACS images at 100 and 160~$\mu$m on 29 July 2011,  a PACS image at 70$\mu$m (and 160$\mu$m)
 on 1 August 2011, as well as  SPIRE images at 250, 350 and 500~$\mu$m on 30 January
2011.  These observations are
summarised in Table~\ref{tab:obs}.

 \begin{table}[!h]
      \caption[]{Herschel observations of  GJ~581. }
         \label{tab:obs}
            \begin{tabular}{l r l r}
            \hline\hline
            \noalign{\smallskip}
   ObsId    &    Date     &  Instrument  &  Integration \\
            \noalign{\smallskip}
            \hline
            \noalign{\smallskip}
                   1342202568                 &  11 August  2010     &    PACS 100/160   &   890s    \\
                   1342213474                 &  30 January   2011  &   SPIRE 250/350/500  &  185s \\ 
                 1342224948                   &  29 July  2011      &    PACS 100/160   &   7190s    \\
                 1342225104                   &  1 August  2011     &    PACS 70/160    &  3936s    \\
  
            \noalign{\smallskip}
            \hline
           \end{tabular}
\end{table}

\subsubsection{PACS} \label{PACSobs}

The PACS observations used the mini-scan map mode with eight legs of a  $3'$ length, with
a $4''$ separation between legs in a single scan direction at a rate of
20~arcsec~s$^{-1}$, and two scan directions ($70^{\circ}$ and $110^{\circ}$).
These data were reduced using the Herschel Interactive Processing Environment HIPE
\citep{Ott10} version 7 and implement version FM6 of the flux calibration.  The
data were pre-filtered to remove low-frequency $(1/f)$ noise using a box-car filter
with a width of 66~arcsec at 70 and 100~$\mu$m and 102~arcsec at 160~$\mu$m.  This
data filtering results in the source flux density being underestimated by
$\sim 20 \pm 5$\% as discussed in detail by \citet{Kenn12}.
Maps were made from these filtered timelines using the photProject task in HIPE.

The pixel scales in the images presented in Fig~\ref{fig:PACS} were set to 1~arcsec at 70 and
100~$\mu$m, and 2~arcsec at 160~$\mu$m, i.e., smaller than the natural pixel
scales. This  enhanced sampling is possible because of the  high level of redundancy
provided by the scan map mode used but it comes at the cost of correlated noise
between neighbouring pixels.  We have also made images with the natural pixel scales
of 3.2~arcsec at 70 and 100~$\mu$m, and 6.4~arcsec at 160~$\mu$m to evaluate the impact
on the  parameter estimation in our modeling. The noise rms for the  images with
the natural pixel are 0.47mJy/$5.6''$beam at  70 $\mu$m, 0.48mJy/$6.7''$beam at 100~$\mu$m, 
and 0.77mJy/$11.4''$beam at 160 $\mu$m.

\subsubsection{SPIRE}

Follow-up observations were taken on 30 January 2011 with SPIRE (Spectral \& Photometric Imaging REceiver, 
\citet{Grif10}) using the small-map mode, resulting in simultaneous 250, 350 and 500~$\mu$m images.  
The data were reduced using HIPE (version 7.0 build 1931), adopting the natural pixel scale of 
6, 10, 14~arcsec at 250, 350 and 500~$\mu$m respectively. 
The  noise rms are 6.1~mJy/$18.2''$beam,  7.9~mJy/$24.9''$beam, and  8.3~mJy/$36.3''$beam 
at 250, 350 and 500~$\mu$m, respectively, and the image at  250~$\mu$m is shown in \S~\ref{SPIRE_imag}. 
 
\subsection{\bf Ancillary data}

\subsubsection{Spitzer} \label{spitzer}

MIPS 70~$\mu$m observations of GJ~581 (AOR 22317568) were taken on 21 August 2007 (no 24~$\mu$m MIPS were taken) 
and a small measured excess, with the significance $\chi_{70}=(F_{70}^{obs}-F_{70}^{*})/\sigma_{70}
=3.6$ in \citet{Kosp09} and $\chi_{70}=2.2$ 
in \citet{Bryd09} with the same data,  forms a tentative discovery.
We  re-reduced the archival data  using an updated pipeline 
and the flux calibration summarized in \citet{Gord07} providing the new flux density
20.0$\pm$5.3~mJy by PSF fitting to the image at the effective wavelength of 71.42~$\mu$m 
(color correction for T$_{dust}$=40~K applied : 0.89 \footnote{http://irsa.ipac.caltech.edu/data/SPITZER/docs/mips/ \break mipsinstrumenthandbook/51/}). 
The uncertainty includes both statistical 
and calibration uncertainties and  the corresponding excess ratio $\chi_{70}$ is 2.7
with our estimate of the photospheric flux density of 5.6~mJy at 71.42~$\mu$m (see \S~\ref{sed}).
Photospheric flux densities predicted for late type stars (K and M) by the Kurucz or Next~Gen models 
 have been shown to be overestimated  in the mid-IR by as much as 5-10\%  
\citep{Gaut07, Lawl09}. Hence, this excess can be treated as a lower limit.

The IRS observations  of GJ~581 (AOR22290432)  were taken on  31 August 2007,  
and the details of the data reduction are  in  \citet{Beic06}  and  \citet{Dods11}.

\subsubsection{HST/NICMOS}

GJ~581 was directly imaged with HST/NICMOS on 6 May 1998 (GO-7894; PI Todd Henry).
The NICMOS data and the overall observing program are described in
 \citet{krist98} and \citet{golimowski04}.
We reanalyzed the F110W data for GJ~581 consisting of 128 seconds
of cumulative integration on the NIC2 camera (0.076$''$/pixel,
256$\times$256 pixels).  Target stars were not placed behind
the occulting spot, near-contemporaneous observations of
PSF reference stars were not made, and multiple telescope roll angles
were not employed. Therefore the observations were not optimized
for high-contrast imaging of low surface brightness circumstellar
nebulosity. Nevertheless, we subtract the GJ~581
point-spread-function (PSF) using observations of LHS~1876 (GJ~250B)
made on  24 March  1998  as part of the same scientific program, and in so doing, set 
constraints on the scattered light disk brightness as discussed in 
\S \ref{fig:nicmos}. 
PSF subtraction techniques, including a discussion of scattered light
artifacts and other spurious features, are described in greater
detailed by \citet{krist98}.

\section{Stellar parameters of GJ~581}  \label{par}

GJ~581 (HIP~74995) lies relatively nearby ($6.338 \pm 0.071$~pc ;  \citet{Phil10}) and is 
classified as a star of spectral type M3.0 \citep{Reid95}.
Recent CHARA interferometric measurements of  its physical radius  ($0.299 \pm 0.010 R_\odot$) 
imply an effective surface temperature of $T_{eff} = 3498 \pm 56 K$, a bolometric luminosity of 
$0.01205 \pm 0.00024L_{\odot}$  and a stellar mass of $0.28 M_{\odot}$  \citep{vBra11}.

\medskip

A variety of different techniques have been discussed in the literature 
as a means to determine the age of GJ~581, including kinematics, magnetic
activity (X-ray observations), chromospheric activity,  stellar color, metallicity and rotation. 
\citet{Legg92} finds that the  galactic velocities of GJ~581  
are intermediate between those typical of the young and old galactic disk M-stars. 
\cite{Bonf05}  conclude that  the low limit on its X-ray emission, 
the low $v~sini$  and
the weak CaII H and K emission,  taken altogether, suggest that GJ~581 is at least 2~Gyr old.
\citet{Sels07} established an $L_x/L_{bol}$ versus age relation for M- K- G- spectral type
stars 
to estimate that the age of GJ~581 could be around 7~Gyr.   
Recently, \citet{Engl11} have established an 
age-rotation period relation for M-stars and determined an age of 
$5.7 \pm 0.8$ Gyr for GJ~581.
Clearly, GJ~581 is an old star well above 1~Gyr.

{High contrast imaging for GJ~581  has revealed no companion of  $\sim$7 Jupiter masses or higher between ~3-30 AU \citep{Tann10}. In addition, the limits provided by the HARPS radial velocity measurements  exclude 
planets that are more massive than  Jupiter  with semimajor axes inside 6~AU \citep[Fig~13 in ][]{Bonf11}.}

All the parameters of GJ~581 are summarized in Table \ref{tab:star_parameter}.

 \begin{table}[!h]
      \caption[]{Stellar parameters of  GJ~581. }
         \label{tab:star_parameter}
            \begin{tabular}{l l l}
            \hline\hline
            \noalign{\smallskip}
   Parameters    &    values     &  references  \\
            \noalign{\smallskip}
            \hline
            \noalign{\smallskip}
    R.-A. ICRS(2000)             &  $\rm 15^H 19^m 27.509^s$           &  \citet{Hog00}    \\
    Dec.     ICRS(2000)          &  $\rm -07^{\circ} 43' 19.44''$      &    ~~~~~''  \\
    $\mu_{\alpha} cos \delta$    &  $-1.228''$/yr                      &    ~~~~~''    \\ 
    $\mu_{\delta}$               &  $-0.098''$/yr                       &    ~~~~~''   \\
    Galactic longitude           &  $354.08^{\circ}$                   &   ~~~~~''  \\
    Galactic latitude            &  $+40.01^{\circ}$                   &   ~~~~~''   \\
    Distance                     &  $6.338 \pm 0.071$ pc               &  \citet{Phil10}     \\
    Spectral type                &  M3.0                               &   \citet{Reid95}   \\
    Radius                       &  $0.299 \pm 0.010 \rm ~R_\odot$     &    \citet{vBra11}  \\
    Mass                         &  0.28 $\rm M_{\odot}$               &   ~~~~~''     \\     
    Bolometric lum.        &  $1.22 \pm 0.02~10^{-2} \rm ~L_{\odot}$  &  ~~~~~''      \\
    Effective temp.        & $ 3498 \pm 56 \rm ~K$               &    ~~~~~''    \\ 
    Metallicity   [Fe/H]         &   $-0.25$                           &   \citet{Bonf05}    \\
    $v~sini$                    & $< 2.1 \rm ~km/s$                   &   \citet{Delf98} \\
    Rotation period              &   $93.2  \pm 1$ days                &   \citet{Vogt10}    \\
    Log$\rm L_X$ (ergs/s)  &    $ < 26.44$                       &   \citet{Schm95}  \\
    Age                          &    2 - 8 Gyr                        &   see \S~\ref{par} \\
            \noalign{\smallskip}
            \hline
           \end{tabular}
\end{table}

\section{Herschel images of GJ~581 } \label{images}

\begin{figure*}[t!]
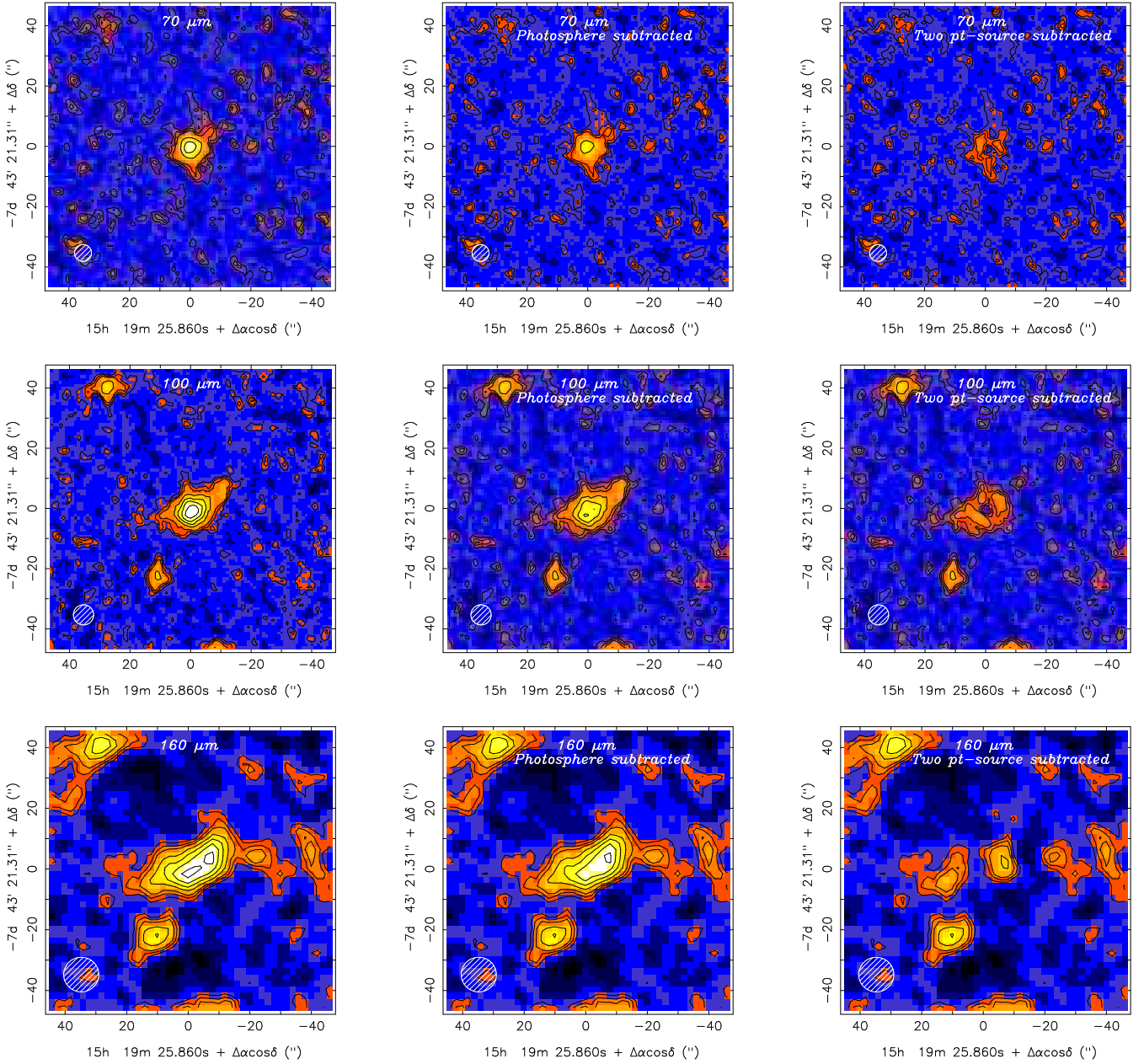

\resizebox{18cm}{!}{\includegraphics[angle=0]{GJ581_70_1.ps} \hspace{2.cm } \includegraphics[angle=0]{GJ581_70_2.ps}  \hspace{2.cm } \includegraphics[angle=0]{GJ581_70_3.ps}  }
\resizebox{18cm}{!}{\includegraphics[angle=0]{GJ581_100_1.ps}  \hspace{2.cm } \includegraphics[angle=0]{GJ581_100_2.ps}  \hspace{2.cm } \includegraphics[angle=0]{GJ581_100_3.ps}  }

\resizebox{18cm}{!}{\includegraphics[angle=0]{GJ581_160_1.ps}  \hspace{2.cm } \includegraphics[angle=0]{GJ581_160_2.ps}  \hspace{2.cm } \includegraphics[angle=0]{GJ581_160_3.ps}  }

\caption{PACS images of GJ~581 cropped to $\pm$50~arcsec from the star, at 70, 100 and 160~$\mu$m from {\it top to bottom}, respectively. 
In the {\it left-hand column}, the raw images show that the main  emission is centrally located about the star
position (image center) and that there are  several point-sources in the field, barely detected  at 70~$\mu$m, 
significantly  at 100~$\mu$m, and more prominently at 160~$\mu$m, as expected for submillimeter background galaxies. 
In \S \ref{images}, we show that the  main emission is extended and centered on the star position, as expected for a debris disk, 
and mingles with a background source $\sim$11~arcsec toward the North West. 
The panels in the {\it middle column} are the photosphere-subtracted images. The panels in the {\it right-hand column} show 
the best subtraction (lowest residuals) of a two-point source model, which assumes that there is no debris disk around GJ~581 
but an extra background source located exactly behind the star in  addition to  the N~W~source.
This model is rejected because of the systematic residuals left, indicative of an extended structure, especially at 70 and 100~$\mu$m.
At each wavelength, the contours levels of the three images are the same and  correspond to  
1,2,3,9,15$\sigma_0$ ($\sigma_0=0.0135$~mJy/$1''$pixel) at 70~$\mu$m, 1,2,3,6,9,12,15$\sigma_0$ ($\sigma_0=0.0094$~mJy/$1''$pixel) at 100~$\mu$m, 
and 1,2,3,5,7,9,11$\sigma_0$ ($\sigma_0=0.0251$~mJy/$2''$pixel) at 160~$\mu$m.
The coordinates of the image center provided in the  labels correspond to  the star position at epoch of observation 
(July 29th -  August 1st 2011). 
The hatched circles are the beam FWHMs  : 5.6, 6.8, and 11.4 arcsec at 70, 100, and 160~$\mu$m, respectively. } 
\label{fig:PACS}
\end{figure*}

In Fig~\ref{fig:PACS}, we present our deep PACS images  of GJ~581 cropped to the region $\pm 50''$ from the star. 
At each wavelength,   
the main emission is close to the star position (image center)  
and the surrounding field is contaminated by several other sources, 
detected  barely  at 70~$\mu$m, significantly at 100~$\mu$m, and prominently at 160~$\mu$m, 
as is expected for submillimeter galaxies in the background. 
The central  emission is suggestive of a spatially resolved  debris disk not fully separated from 
the background source to the N~W.
In the next subsections, we analyse quantitatively these PACS images to verify this view.



\begin{figure*}[t!]
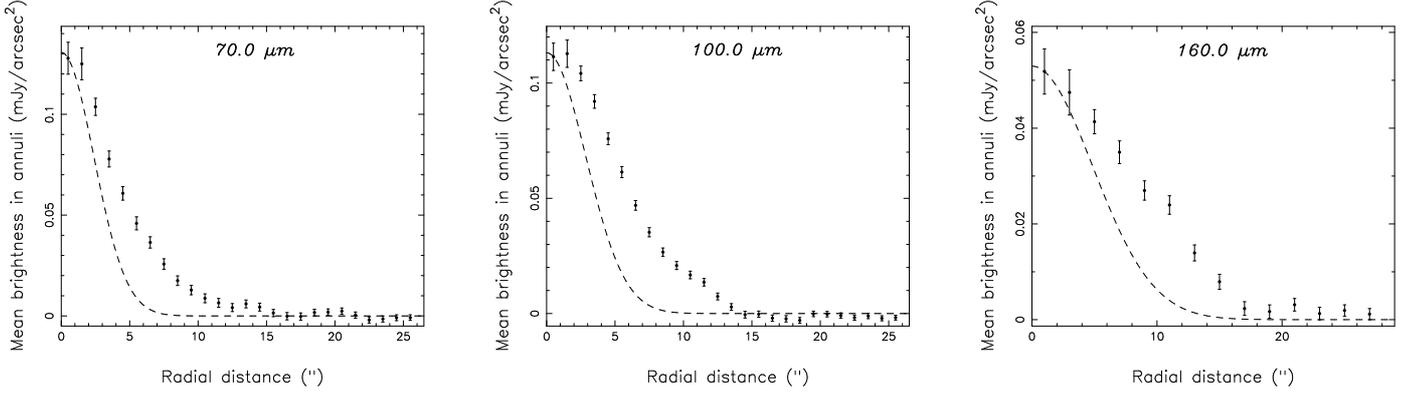

\resizebox{18.5cm}{!}{\includegraphics[angle=0]{GJ581_70_prof_40deg.ps} \hspace{2.cm } \includegraphics[angle=0]{GJ581_100_prof_40deg.ps}  \hspace{2.cm } \includegraphics[angle=0]{GJ581_160_prof_40deg.ps}  }
\caption{ {Radial profiles of the mean brightness of the photosphere-subtracted images. Each point of these curves 
was computed as the mean of the emission in  an elliptical annulus centered on the peak 
of the main emission close to the image center. They correspond to an axisymmetric disk model inclined to $40^{\circ}$
 relative to the plane of the sky (see text).  They are one pixel wide, {\it i.e.}  1 arcsec in the 70 and 
100~$\mu$m images and 2 arcsec in the 160~$\mu$m image.  The Gaussian with the FWHM of the beam is the  profile expected for 
a hypothetical point source in the background aligned by chance with the star and shown for comparison.
} }
\label{fig:prof}
\end{figure*}

\subsection{Radial profiles of the emission in the PACS images}

First, in Fig~\ref{fig:prof}, we present the radial profiles of the  emission 
at the three wavelengths by computing the mean brightness in ellipical annuli centered 
on the peak of the main emission. At the three wavelengths, these radial profiles show 
that the  emission is  extended about this peak  when they are compared
to the  Gaussian profiles of a hypothetical point source in the background.

We computed these profiles after subtracting from each image a PSF scaled to the  photosphere flux density 
($S_{photosphere}=5.8, 2.8, 1.1$~mJy at 70, 100 and 160~$\mu$m, respectively, see \S \ref{sed}). 
The emission peak position was found to be less than 2~arcsec  from the star in each image, 
consistent with the pointing accuracy of  the {\it Herschel} telescope \footnote{http://herschel.esac.esa.int/twiki/bin/view/Public/SummaryPointing}. 
We have tested elliptical  annuli at PA =120$^{\circ}$ and with inclinations $0^{\circ}$ (circular), 
$25^{\circ}$, $40^{\circ}$ and  $75^{\circ}$, anticipating that the disk may  not be in the plane of the sky. 
We found that all these radial profiles were  more extended than the Gaussians  at the three wavelengths, and
slightly more for the inclination  of $40^{\circ}$.  
The imprint of the N~W source at the radial distance of 11~arcsec can be seen in these profiles 
at 100 and 160~$\mu$m. We have tested the method by computing the radial profile of the emission of the 
South East background source in the same way. We satisfactorily found that its profile matches within $1\sigma$  the
Gaussian  expected for a point source.

The extended emission revealedby these profiles can also be seen directly in the photosphere-subtracted 
images of Fig~\ref{fig:PACS} (middle column), 
most prominently at 70~$\mu$m because the photospheric flux density is highest at this wavelength.

\subsection{Gaussian source fits } \label{Gaussfit}


As a second approach to verify that the central emission is more extended than the PSF, we fit an elliptical 2D Gaussian to each photosphere-subtracted image with masking applied to the position of the N~W~source at 100 and 160~$\mu$m.
At 70~$\mu$m, we found FWHM of the minor and major axes of $9.5\pm0.7''$ and $10.7\pm1.1''$; a position angle of$120 \pm10^{\circ}$; and a flux density of $18.8\pm1.4$ mJy after adding back in the photospheric contribution.
At 100~$\mu$m, we found, respectively, values of $9.9\pm1.0''$, $13.3\pm0.5''$,  $120\pm10^{\circ}$,
and $21.1\pm1.5$~mJy  for the minor and major axes FWHM, the position angle, and the 100~$\mu$m flux density,
after masking the N~W~source (all pixels in a $12'' \times 12''$ square centered at $-9''$ and $+6''$ 
from the star).
Given that the FWHM of the PACS PSFs are $5.6''$ and $6.8''$ at these wavelengths, 
we conclude that the emission is significantly extended, as already shown by the radial profiles,  
and is elongated at PA $\sim 120^{\circ}$.

The ICRS coordinates of the  70~$\mu$m Gaussian peak in this fit  are 15$\rm^h$~10$\rm^m$~25.905$\pm0.05\rm^s$ 
and $-7^{\circ}~43'~22.79\pm0.7''$ and differ only by $0.3''$ from the adjusted position of the 100~$\mu$m Gaussian peak.
These coordinates differ by $+0.66''$  and $-1.48''$ from the right ascension and declination
of the star GJ~581 predicted with the {\it Hipparcos} astrometric parameters 
(Table~\ref{tab:star_parameter}). These differences are consistent
with the  $1\sigma$  pointing accuracy of $2''$ for the {\it Herschel} telescope.
We conclude that the main emission in the PACS images at 70 and 100~$\mu$m 
is centered on the star position within pointing uncertainty.

The 160~$\mu$m image is the product of the coaddition of 
two images taken independently with the PACS 70/160 and PACS 100/160 instruments 
only a few days apart in 2011 (Table~\ref{tab:obs}). 
The registration of these two images were facilitated 
by the negligible displacement due to proper motion over the short lapse of time and by
the  fortuitously small  difference of $0.3''$  between the pointing positions of 
the two instruments as found above. 
Our first 160~$\mu$m image  in 2010 was not coadded because no feature in the image could be used
to check the registration since its signal-to-noise ratio was $\sqrt{12.5}$ times lower. Although this
first image  was crucial in our decision to observe deeper, its use or not is inconsequential
for our analysis. For the fit at 160~$\mu$m, 
we fixed the position of the Gaussian to the coordinates 
determined at 70~$\mu$m. This was necessary because the large mask ($16'' \times 16''$ square) used for the N~W source affected the independent determination of this position. The best fit 2D Gaussian parameters were minor and major axes of $12.8\pm1.5''$ and $21.5\pm2''$; a position angle of $125 \pm10^{\circ}$; and a flux density of $22.1 \pm 5.0$~mJy, after adding back in the photospheric contribution.  Given the PACS PSF at 160~$\mu$m of $11.4''$, this indicates that the emission is  extended  at this wavelength as well. 

The disk inclinations resulting from the ratios of the minor  and major axes determined above, 
and corrected quadratically for  the convolved PSFs,  are  :  
$33 \pm 17^{\circ}$ ($0^{\circ}$ is face-on),  $54 \pm 6^{\circ}$,
and  $71  \pm 7^{\circ}$ at  70, 100 and 160~$\mu$m respectively. Although scattered, these values are
statistically consistent, since they are within  $1.5\sigma$ from their weighted mean (59$^{\circ}$), 
and indicate an inclined disk which has implications for the masses of the planets of the system as 
discussed in \S~\ref{planets}.

We note that the three major axes above, corrected quadratically for the convolved PSF, 
are very closely proportional to the wavelength,
suggesting that the disk is radially broad  since emission at longer wavelength 
probes colder dust, more distant from the central star. 

The flux densities from our fits above have been scaled up to account for the flux removed by the 
data filtering during the reconstruction of the images; 
the correction factors are 16, 19 and 21\% with an uncertainty of 5\% estimated for point sources
in the DEBRIS survey by \citet{Kenn12}.   
The uncertainty of each PACS flux density determined above is based on the quadratic sum of the statistical uncertainty 
in our Gaussian fit,  the absolute flux calibration accuracy of 3\% at 70 and 100~$\mu$m, and 5\% 
at 160~$\mu$m, provided by the {\it Herschel} project 
\footnote{http://herschel.esac.esa.int/twiki/pub/Public/PacsCalibrationWeb/}, 
and  the 5\% uncertainty of the correction factor for the data filtering. The uncertainty of the flux
density at 160~$\mu$m is formally 2~mJy with this calculation but our fit  depends 
to some degree on the position of the mask applied for the N~W background source. So we
 have increased this uncertainty to 5~mJy at this wavelength based on several fits with different masks.

\subsection{Considering the superposition of two point-sources}

As a final test, 
we consider the possibility that the extended emission in the central part of the image
could be caused by the superposition of two backgound sources instead of a disk. 
To this end, we subtracted two PSFs from each PACS  image (in addition to subtracting the phostospheric
emission),  adjusting their flux densities 
in order to remove as much emission as possible. The first PSF was located at the position 
of the N~W~source, {\it i.e.} at $-9''$ and $+6''$ from the star, and  the second was tested at six locations ;   
the star position itself,  as well as a half FWHM to the North, South, East and West of the star, 
and half way between the  star and the N~W source. The lowest residuals were found after removing  
1.4, 2.9 and 6.6~mJy for the  first PSF at 70, 100 and 160~$\mu$m, respectively, and  
4.9, 6.9 and 9.4~mJy for the second PSF at the star position at 70, 100 and 160~$\mu$m, respectively.
Note that these latter flux densities are free of the photosphere contributions estimated in \S~\ref{photosphereflx}.
Despite this removal process, there is still significant structure left in the residual images  
shown as the {\it two-point source subtracted images} in Fig~\ref{fig:PACS} (right-hand column). 
This structure can be best explained as resulting from the 
extended emission of the disk incompletely removed by this process. Hence, 
we conclude that this test rejects the possibility that the superposition of two background 
point sources can be responsible for the  central emission.

We elaborate further by discussing the probability  to find such contaminant sources 
in the field and their spectra.
First, the probability to have one background source stronger than 6.6~mJy  at 160$~\mu$m within $11''$ 
from the star  is  18 \%,  and to have two is  only 1.8\% by using the Poisson probability distribution 
with the mean source surface density N($S > 6.6$mJy) $\sim$ 4000 sources/deg$^2$ at $160\mu$m  
provided by the Herschel PEP survey \citep{Bert11} (see also \citet{Sibt12}). 
Second, the spectra of these test sources removed from the images may or may not be physical.
The flux densities removed at the N~W~source position
(1.4, 2.9 and 6.6~mJy at 70, 100 and 160~$\mu$m respectively)
 are consistent with the spectrum of a galaxy at $z>1.5$ according to  Fig~4 of \citet{Blai02} valid 
 for a typical high-z  galaxy enshrouded in dust ($\sim$ 38~K, $L \sim 5 \times 10^{12} L_{\odot}$)
and radiating in the far-IR and submm.
However, the flux densities removed at the star position  
(4.9, 6.9 and 9.4~mJy at 70, 100 and 160~$\mu$m respectively)  above the photospheric levels  
make the ratio $S_{100\mu m}/S_{70\mu m}$  lower than expected for a galactic spectrum according to that work.

\subsection{SPIRE images}  \label{SPIRE_imag}

\begin{figure}[h!]
\centering
\resizebox{5.54cm}{!}{\includegraphics[angle=0] {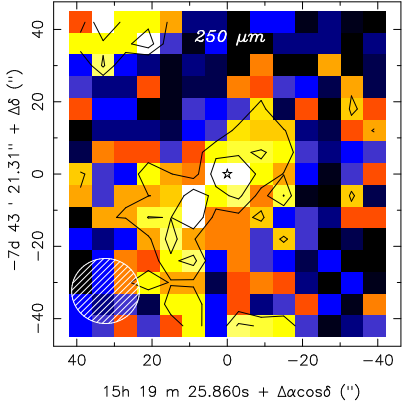}}
\caption{SPIRE image of GJ581 at 250~$\mu$m.  Pixel size is $6''$ and the contour levels are 1 
and $2\sigma$ with $\sigma$= 6.1~mJy/$18.2''$beam. The star symbol is the position of GJ~581
at the date of observation.} 
\label{fig:spire}
\end{figure}

The SPIRE image at 250 $\mu$m in Fig~\ref{fig:spire} shows an elongated structure 
at PA $\sim 120^{\circ}$ which has two  peaks at the $2\sigma$ level that match the positions of 
the star and  the N~W~source. This structure is also extended to the S~E. 
Although this structure is of low statistical significance, it is consistent with the emission 
detected at the PACS wavelengths. The two other SPIRE images, at 350 and 500~$\mu$m, are dominated by noise
and no reliable structure can be recognized. 
A 2D~Gaussian could not satisfactorily model the emission at 250~$\mu$m, and so we
carried out photometry with an aperture of 36~arcsec and measured a flux density of $24 \pm 6$ mJy. 
This flux density is considered an upper limit for the disk since its emission is blended 
with that of the N~W~source.
  
\subsection{Spitzer MIPS image at 70 $\mu$m}

A {\it Spitzer} MIPS image at 70~$\mu$m was made  on 21 August 2007,
four years before our PACS image (1 August 2011).
 We fit a Gaussian with the FWHM of 19~arcsec (70~$\mu$m MIPS beam)
to the emission of the MIPS image and found  the  
coordinates of the Gaussian peak to be at  $\alpha=15\rm^h$~10$\rm^m$~26.08$\pm0.17\rm^s$ and 
$\delta=-7^{\circ}~43'~23.2\pm2.5''$ in the ICRS system. 
The relatively high  noise of the MIPS image did not permit a solution for the FWHM.
The differences in the coordinates compared with our 70~$\mu$m PACS position 
given in \S~\ref{Gaussfit} are  $\Delta \alpha=-2.56''$ and $\Delta \delta=+0.45''$ (PACS minus MIPS)
with an uncertainty of $3.3''$ when combining the astrometric uncertainty of the MIPS Gaussian fit  
($2.5''$ in both coordinates)  and  the  pointing accuracies of $2''$ for Herschel and of $1''$ for 
Spitzer \footnote{http://irsa.ipac.caltech.edu/data/SPITZER/docs/spitzermission/ \break missionoverview/spitzertelescopehandbook/12/}. 
The predicted displacement of GJ~581 between the two epochs of observations is $\Delta \alpha=-4.98''$ 
and $\Delta \delta=-0.41''$ computed with the proper motion in Table~\ref{tab:star_parameter}.
Hence, the coordinate differences of the 70~$\mu$m  emission measured between the two epochs are compatible with this
prediction but the star has not moved sufficiently between these epochs for us to 
confirm that the 70micron emission is comoving with the star at a 
statistically significant level.

\begin{table*}[t!]
\caption[]{Best fit models of the disk for each individual image and for the combined fit.}
\label{tab:bestfits}
\begin{tabular}{l c r c r c c c c c}
\hline\hline
\noalign{\smallskip}
 image(s)    &  $\chi_{\nu}^2$  &  ~~~$\nu~~~^a$   & $ A~~~^b$        &  $r_{inner}$        & $r_{outer}~^c$           & $\alpha$  & $f_T$      &       $i~^d$     &    $\Omega~^e$      \\
             &                  &             & ($AU^2$)    &    (AU)             &   (AU)                &           &            &  $(^{\circ})$ &  $(^{\circ})$    \\
             &                  &             &                   &                     & lower~~(formal)       &           &            &               &                  \\
            \noalign{\smallskip}
            \hline
             &                  &             &                   &                     &                        &                            &                &              &                 \\

 70 $\mu$m   &      1.04        &  408        &     $0.8 \pm 0.5 $        & $25_{-14}^{+20}$     &  $\ge 40$ ~~~~~~(62)   &  $-2  \le \alpha  \le 0 $  & $4.0_{-1}^{+0.5}$ &    $ \le 60$ & $120 \pm 20$    \\
             &                  &             &                   &                     &                      &                            &                &              &                 \\

100 $\mu$m   &      0.99        &  408        &  $ 2.2 \pm 1.2 $          &  $31_{-14}^{+22}$  &  $\ge 55$ ~~~~(100)      &   $-2  \le \alpha  \le  0 $  &  $3.5_{-1}^{+0.5}$  &    $ \le 60$ & $110 \pm 20$    \\
             &                  &             &                   &                     &                       &                            &                &              &                 \\
 
160 $\mu$m   &      0.97        &   90        &  $ 1.5 \pm 0.7 $          & $37_{-12}^{+18}$    &  $\ge 90$ ~~~~(145)    &   $-2  \le \alpha  \le 0 $  &  $4.5_{-1}^{+0.5}$  &    $ 40 < i <  80$ & $120 \pm 20$    \\
             &                  &             &                   &                     &                       &                            &                &              &                 \\

Combined     &       1.03       &   922       &  $ 2.3 \pm 1.1 $         &  $ 25_{-12}^{+12}$  &  $\ge 60$ ~~~~(110)  &   $-2  \le \alpha  \le 0 $  &  $3.5_{-1}^{+0.5}$  &   $ 30 < i < 70$ & $120 \pm 20$    \\
             &                  &             &                   &                     &                       &                            &                &              &                 \\
\noalign{\smallskip}
\hline
\end{tabular}

{\small 
$^a$ : based on the number of natural pixels (the natural pixel is $\sim$ 3 times larger than the pixel in the images shown).\filbreak 
$^b$ : $A$ is the total cross-sectional area of all the grains (see \S~\ref{Model}) \filbreak 
$^c$ : lower limit (see text)  and, in parenthesis, the formal value corresponding to the minimum  $\chi_{\nu}^2$ given. \filbreak
$^d$ : $i=0^{\circ}$ is face on. \filbreak
$^e$ : $\Omega > 0$ is E of N }
\end{table*}

\section{ Modeling the PACS images of GJ~581} \label{mod}
 
\subsection{Parametrized model }   \label{Model}
   
We fit a parametrized model of the disk to the PACS images. 
The model is axisymmetric and truncated by the inner radius $r_{in}$ and the outer radius $r_{out}$ 
which  are free parameters in our fit.
Its dust emission is optically thin, and the flux density from each element~$(k,l)$ of the grid 
covering the disk is

$$ \Delta S_{k,l}=  \epsilon B_\nu(T(r_{k,l})) ~.~ \Sigma_p  r_{k,l}^{\alpha} ~.~ \Delta a   /  d^2, ~~~~~(1)$$

\noindent where $B_\nu(T(r_{k,l}))$  is the Planck function  that depends on the
grain temperature $T(r_{k,l})$ at the radial distance $r_{k,l}$ from the star,   
$\Delta a$ is the area of the element in the grid, $d$ is the distance to the star,
and  $\Sigma_p$ is the coefficient of the power-law 
\citep[e.g.][]{Wyat99}~\footnote{There is a different but equivalent derivation of the flux density given 
in \citet{Zuck01} based on the surface emittance $\pi B_{\nu}$.}. 
To fit individually each PACS image in \S~\ref{individual}, we set the factor $\epsilon$  to unity in eq~(1).
To fit simultaneously the three PACS images at $\lambda=70~\mu$m,
100~$\mu$m, and  160~$\mu$m in \S~\ref{combined}, we implement a grey body effect \citep[e.g.][]{Dent00} 
 by setting  $\epsilon$ to unity if $\lambda < \lambda_0$,  and  
 $\epsilon = 1.0 \times (\lambda_0 / \lambda)^{\beta}$  if $\lambda > \lambda_0$,
where  $\lambda_0$ and $\beta(> 0)$ are free parameters in our fit. 

In our model, no assumption is made for the size distribution of the grains, 
their mineralogical composition and  porosity. The thermal structure  of the disk is taken as 

$$T(r)=f_T ~.~ T_{BB} (r),  ~~~~~(2)$$ 

\noindent where  $f_T$  is a scaling factor applied to the black body temperature
$T_{BB}(r)=278 ~.~ (L/L_{\odot})^{0.25} ~.~ (r/1 AU)^{-0.5}~(K)$, and is a free parameter in our fit.
Here we illustrate how to interpret this parameter using
a simplified model of the absorption and reemission of the starlight by the grains. 
For dust with the absorption and emission efficiencies $Q_{abs}$ and $Q$, a straightforward   
derivation shows that  $f_T=(Q_{abs}/Q)^{1/4}$  for a grain at thermal equilibrium, ignoring that these
 efficiencies should be averaged over the spectrum of 
incoming and outgoing radiation, and integrated over the dust size 
distribution. If we assume that grains larger than 
1~$\mu$m absorb starlight with the efficiency $Q_{abs}=1$
and reemit it at longer wavelengths at the lower efficiency $Q$,  then
the simple interpretation is that $f_T$ is directly related to the emission efficiency  
through the  relationship $f_T=Q^{-1/4}$. 
  
The  term $\Sigma_p  r^{\alpha}$  in eq~(1) is the emitting cross-sectional area 
of the grains per unit area of the disk surface.
These  grains are spatially distributed according to a radial profile taken 
as the power-law $\Sigma_p  r^{\alpha}$, and their  total cross-sectional area is $A$. 
Since these grains reemit with the efficiency $Q$, their total emission 
is proportional to  $Q~.~A=\int_{r_{in}}^{r_{out}}~ 2\pi r dr \Sigma_p  r^{\alpha}$.   
Hence, if $\alpha \ne -2$, the coefficient of the power-law is 

$$\Sigma_p=  f_T^{-4} ~.~ A ~.~ (\alpha+2)~/~(2\pi~(r_{out}^{\alpha+2}-r_{in}^{\alpha+2})), ~~~~~(3)$$ 

\noindent and the total cross-sectional area $A$ and the power-law index $\alpha$ are  free parameters
in our fit \footnote{if $\alpha=-2$,
$\Sigma_p= f_T^{-4}  ~.~ A ~/~(2\pi~(ln(r_{out})-ln(r_{in})))$}.
Grains smaller than 1~$\mu$m are not important because they emit so inefficiently that their flux density 
is negligible \citep{Bons10}. 

The  total cross-sectional area $A$ can be converted into  mass  assuming 
a  size distribution and a mass  density $\rho$  for the material.  
Adopting the standard size distribution  $n(D) \propto D^{-3.5}$ for spherical particles of diameter $D$
between $D_{min}$ and  $D_{max}$,  the corresponding mass is
 
$$m_d=(2/3) ~.~ A  ~.~ \rho  ~.~ \sqrt{D_{min}}   ~.~ \sqrt{D_{max}}. ~~~~~(4)$$

This model is complemented with a point source photosphere 
centered on the image by two free parameters (coordinates $x_c$ and $y_c$) and having  
the flux densities estimated  from the Next~Gen stellar 
atmosphere model in \S~\ref{sed} but lowered  by $\sim20\%$ because of the data filtering used 
to reconstruct the images as already mentioned in \S~\ref{PACSobs}.  
This model is projected
onto the sky with the inclination $i$ and the node orientation $\Omega$,  
and finally convolved by the telescope PSF  provided by the PACS images of the reference star $\alpha$~Boo 
at 70 $\mu$m, 100 $\mu$m and 160 $\mu$m. 
Hence, our model has 9 free parameters ($r_{in}, r_{out}, \alpha, f_T, A, i, \Omega, x_c, y_c$). 

The best fit  is found by minimizing  

$$\chi_{\nu}^2 = \sum_k  \sum_l{ \Big( { O_{k,l}-C_{k,l} \over \sigma_0} \Big)^2}, ~~~~~(5)$$  

\noindent computed with the residuals between the image (O) and the model (C) over all the pixels of the image, and assuming the same  
measurement uncertainty $\sigma_0$ for all the pixels. To compute the model,  
we used a grid  on the sky which has a resolution  of 0.5'' at 100 $\mu$m and 70 $\mu$m, and 1'' at 160 $\mu$m, {\it i.e.}
twice as fine as the pixel size of the images of Fig~\ref{fig:PACS}, 
and has dimensions $128 \times 128$ at 70 and 100~$\mu$m, 
and $64 \times 64$ at 160~$\mu$m. These dimensions can accommodate  the largest disk model tested
($r_{out}$=150AU) extended by  twice the beam FWHM. This sky grid is the same
for all the models tested so that the number of degrees of freedom ${\nu}$ is the same for all of them. 
To use the $\chi^2_{\nu}$ probability distribution to discriminate between them, 
we carefully estimate the {\it a priori} uncertainty $\sigma_0$ 
by computing the noise rms over the image limited to the sky grid dimensions
and after excluding the central  part ($ r < 25''$) where the disk emission is located.
For GJ~581, the noise rms, $\sigma_0$,
is 0.0135mJy/$1''$pixel at 70~$\mu$m,  0.0094mJy/$1''$pixel at 100~$\mu$m,  
and 0.0251mJy/$2''$pixel at 160~$\mu$m, corresponding to 
0.49mJy/$5.6''$beam, 0.50mJy/$6.7''$beam, and 0.81mJy/$11.4''$beam, respectively, by
using the  beam area $\pi \times \rm{FWHM}^2 /4\ln2$.

Finally, we used the  SPIRE $3\sigma$ upper limits as a constraint to  reject any model with
a flux density of the dust emission larger than 24~mJy at 250~$\mu$m.

\begin{figure*}[t!]
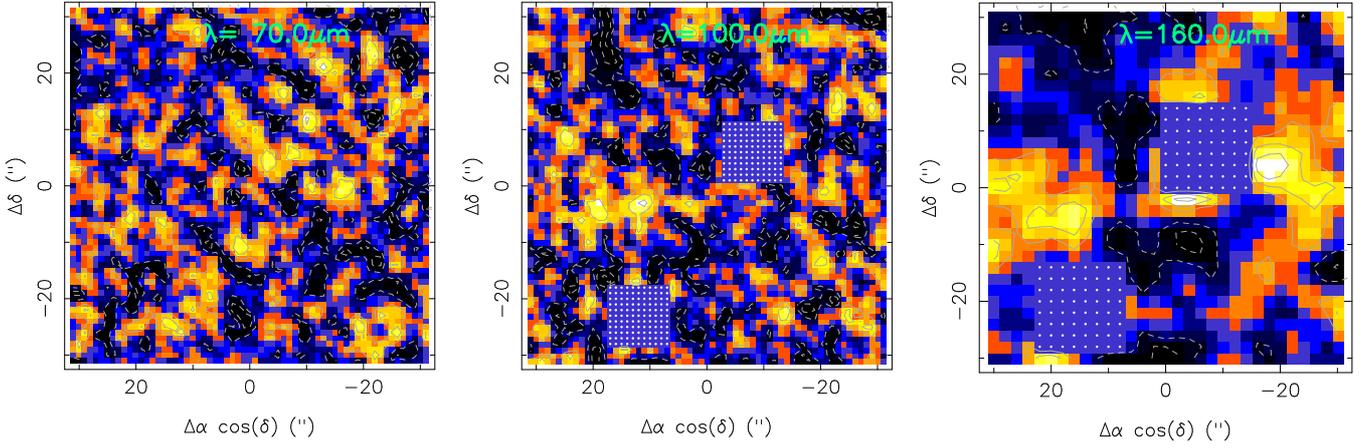

\resizebox{18cm}{!}{\includegraphics[angle=-90]{GJ581_70_res.ps}  \includegraphics[angle=-90]{GJ581_100_res.ps} \includegraphics[angle=-90]{GJ581_160_res.ps}}
\caption{Maps of residuals for the best fit models of the disk to  
the PACS images at 70, 100, and 160~$\mu$m  {\it from left 
to right}. The post-fit residuals  are between  $\pm 3.5 \sigma_0$, coded as black, blue, orange, yellow 
and white over this range ($\sigma_0$ is the same noise rms  as used for contours in Fig~\ref{fig:PACS}
but the color scale is not the same as in Fig~\ref{fig:PACS}).
In zooming the  electronic version, the  contours apparent in these maps are -3, -2, -1, 1, 2, 3   $\sigma_0$,  
 (dashed contours are negative levels).  
The two background sources are masked in the 100 and 160~$\mu$m images as the dotted squares show. }
\label{fig:residuals}
\end{figure*}

\subsection{Fits of individual PACS images} \label{individual}

First, we searched for the best fit model for each individual image. 
The ranges of the model parameters tested were: $A$ of 1  to 20~AU$^2$;  
$r_{in}$ of 3.1  to  80.0~AU; $r_{out}$ of $r_i$~ to  150.0~AU; $\alpha$ of -3.0 to 0.0;
$f_T$ of 1.0 to  6.0; $i$ of 0.0 to  90.0$^{\circ}$ (0.0$^{\circ}$ is a disk seen face-on); 
and $\Omega$ of  0.0 to  180.0$^{\circ}$ (0.0$^{\circ}$ is North and increasing  $\Omega$ is East).
The range for the index $\alpha$ was chosen to cover
possibilities such as grains being blown out of the system by radiation pressure ($\alpha=-1$), and having a distribution akin to that of the Minimum Mass Solar Nebula (MMSN) ($\alpha=-1.5$), as discussed in the modeling of the disk around
the A-star $\beta$~Leo by \citet{Chur11}.  
The two background sources, N~W and S~E of GJ~581, were 
masked as shown in the residual maps
of Fig \ref{fig:residuals}. Roughly 50 million models were tested in our search for the best fit.   

We found the reduced $\chi^2_{\nu} = $  1.04, 0.99, and 0.97 
for the best fits to the three images at 70, 100, and 160~$\mu$m, respectively.
The numbers of degrees of freedom are $\nu=$~408 at  70 and 100~$\mu$m, and $\nu=$~90 at 160~$\mu$m.
 These reduced  $\chi_{\nu}^2$  indicate noise-like post-fit residuals according to  $\chi^2$-statistics.
The residual maps in Fig~\ref{fig:residuals} do not show any systematic residuals as expected in these
conditions.  We stress that  the uncertainty $\sigma_0$ used for  eq~(5) is 
the value determined {\it a~priori} and is not purposefully tweaked {\it a~posteriori} to make the reduced $\chi^2_{\nu}$  close to unity. 
 
The  best fit values of the parameters are in Table \ref{tab:bestfits}.  
There are significant correlations between parameters, especially between $A$, $\alpha$, $r_{outer}$  
and $f_T$, as we found by inspecting the two-dimensional projections of the $\chi^2_{\nu}$ hypersurface, {\it e.g.} 
Fig~\ref{fig:correl} for the pair $A$ and $f_T$. 
To estimate the parameter uncertainties  in these conditions, we have determined the lower and upper limits around  
the best fit value of each parameter  that correspond to  the  fits in which  the reduced $\chi^2_{\nu}$  
are increased to 1.12 and 1.25 with all the other parameters freely adjusted.
These thresholds correspond to a probability of 5\% in $\chi^2_{\nu}$-statistics that  
the reduced $\chi^2_{\nu}$ of pure noise  exceeds 1.12 and 1.25  for the number of degrees of freedom  $\nu=$~408 and 90, respectively.
This is a standard criterium in fitting procedures. We have also inspected the corresponding residual maps 
and noticed  nascent systematics for these degraded fits as expected.  
For the outer radius $r_{out}$, only the lower limits  and the best fit values of the fits
are given in Table \ref{tab:bestfits} because the upper limits are not well constrained since any distant dust becomes
very cold, even accounting for the incident interstellar radiation field as a source of heating 
\citep[][Fig~A1]{Lest09}. 
The resulting range for $r_{in}$ and $r_{out}$ does not permit a conclusive estimate of the radial breadth of the GJ 581 disk.

\subsection{Combined fit of the three PACS images} \label{combined}

In order to consolidate these results and to break correlations
between parameters, we  combined the three PACS images at  $\lambda=70$, 100, and 160~$\mu$m  in a single fit by 
setting the factor $\epsilon$ of eq.~(1) to  unity if  $\lambda < \lambda_0$ and  
to $\epsilon = 1.0 \times (\lambda_0 / \lambda)^{\beta}$  if  $\lambda > \lambda_0$, 
in order to implement a grey body effect. Our search covered 
successively the combinations of $\beta=0.0,~ 0.5, ~ 1.0, ~ 1.5,  ~ 2.0$ and $\lambda_0=70~\mu$m,  
~$85~\mu$m, $130~\mu$m. The ranges of the other model parameters tested were  the same as 
for the individual images in \S~\ref{individual}. The two background sources were masked. 
Note that the 160~$\mu$m image with 4 times fewer pixels has a lower weight 
than the two other images in this combined fit.

\begin{figure}[h!]
\resizebox{9cm}{!}{\includegraphics[angle=-90]{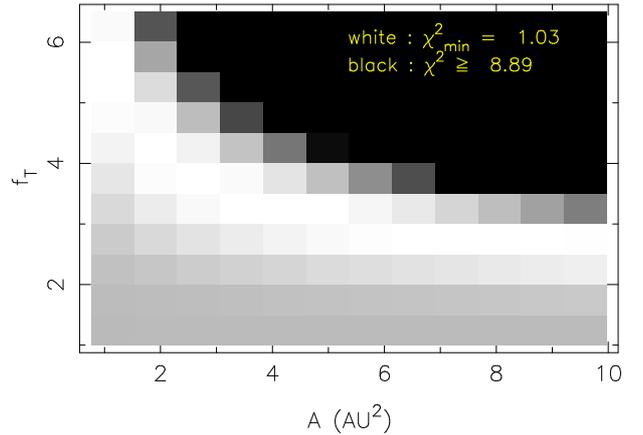}}
\caption{Map of the reduced $\chi^2_{\nu}$ showing the correlation between the temperature factor $f_T$ and 
the total cross-sectional area $A$  of the dust. The minimum $\chi^2_{\nu}$ is  the white region in this map.}
\label{fig:correl}
\end{figure}

The best fit has the reduced $\chi^2_{\nu}$ of 1.03 ($\nu$= 922) 
for $\beta=0$, indicating formally no grey body effect for $\lambda_0 < 160~\mu$m
and so a disk dominated by large grains.
However, there is a high correlation between $\alpha$ and the pair ($\beta$, $\lambda_0$) 
so that these parameters cannot be properly constrained in reality. 
In fact, in the discussion below, we argue that small grains should be abundant in the disk
by comparing timescales of collision and removal processes.   
The best fit values of the other parameters are in Table \ref{tab:bestfits} 
and their uncertainties were determined as described in \S~\ref{individual}.
The total cross-sectional area of the dust $A=2.3$~AU$^2$ can be converted to the dust mass 
$m_d=2.2 \times 10^{-3} \sqrt{D_{\rm{max}}/10~{\rm cm}}$ in $M_\oplus$ for a collisional cascade, using eq(4) 
with $\rho=1.2$~g/cm$^{3}$ for icy grains and $D_{min}=1~\mu$m. The maximum diameter 
$D_{max}$ is unconstrained although objects larger than 10cm contribute negligibly to the
emission at the wavelengths considered in this paper.
Nonetheless, the size distribution probably extends beyond 10~cm as discussed in \S~\ref{timescales}.  
The inclination could be anywhere within a relatively broad range ($30^{\circ} < i < 70^{\circ}$)
that  matches  the purely geometrical determination based of the ratio of the major and minor axes
of the Gaussians fit to the central emission in \S~\ref{Gaussfit}.
The inner radius is $25\pm12$~AU potentially providing an indication of the scale of the planetary system around GJ 581. 
In a similar way to the fits of the individual images, we cannot  distinguish  between a relatively narrow ring
and a disk extending  beyond 100~AU with this combined fit. 
The best fit value of $f_T$ is $3.5^{+0.5}_{-1.0}$, making the  dust temperature
between 50 and 30 K over the extent of the disk, despite the low luminosity of GJ~581.      
This factor is partially correlated with $A$  as shown in Fig~\ref{fig:correl} but it is clearly inconsistent with unity
as we have established by forcing $f_T=1.0$ in the model and found 
a reduced $\chi^2_{\nu}$  as high as 1.92 for the best fit with this constraint.  
An emission model including a grain size distribution instead of a fixed $f_T$ as in our current model   
would be more realistic, but this can be best implemented only if the SED is finely sampled spectroscopically 
from mid-IR to submm  \citep[e.g.][]{Lebr12}.

\subsection{ Model with a Gaussian profile for the grain surface density}

Finally, instead of a power law for the radial distribution of the grain surface density, 
we tested a Gaussian profile $\Sigma_g\exp(-0.5\times((r-r_g)/w_g)^2)$
peaking at radius $r_g$ and having FWHM of  $w_g \times 2\sqrt{2 \ln 2}$. 
 The sky grid for the model, 
the calculation of $\chi^2_{\nu}$, and the ranges of model parameters  
tested  for $A$, $f_T$, $i$, and $\Omega$ were as in \S~\ref{individual}. 
Values of  $r_g$~ ranged from $2 \times w_g$ to 150AU, and $w_g$~ ranged from 3.1AU to $r_g/2$ 
(the Gaussian profile is truncated to $2 w_g$ toward the star). We modeled the three PACS images 
individually and in combination.  The resulting best fits have reduced $\chi^2_{\nu}$ of 1.03,  0.94, and 1.05 
at 70, 100, and 160~$\mu$m, and of 1.04 for the combined images with $\beta=0$ (no grey body effect).
The residual maps are featureless. 
These best fits are  statistically indistinguishable from those with the power law presented in \S~\ref{individual} and \ref{combined}.
The resulting  parameters are :   $r_{g} = 52 \pm 15$AU,  FWHM=  
 $38 \pm 15$AU,~ $A=2.5 \pm 1.2$~AU$^{2}$, ~$f_T=3.0 \pm 0.5$,
~$i < 60^{\circ}$, and $\Omega = 120 \pm 20 ^{\circ}$. 

In this model, the inner part of the system is populated with dust making the inner radius  
determined with the power law surface density in \S~\ref{combined}  less definitive. 

\section{The SED and IRS spectrum of GJ~581} \label{sed}

\begin{figure*}[!t]
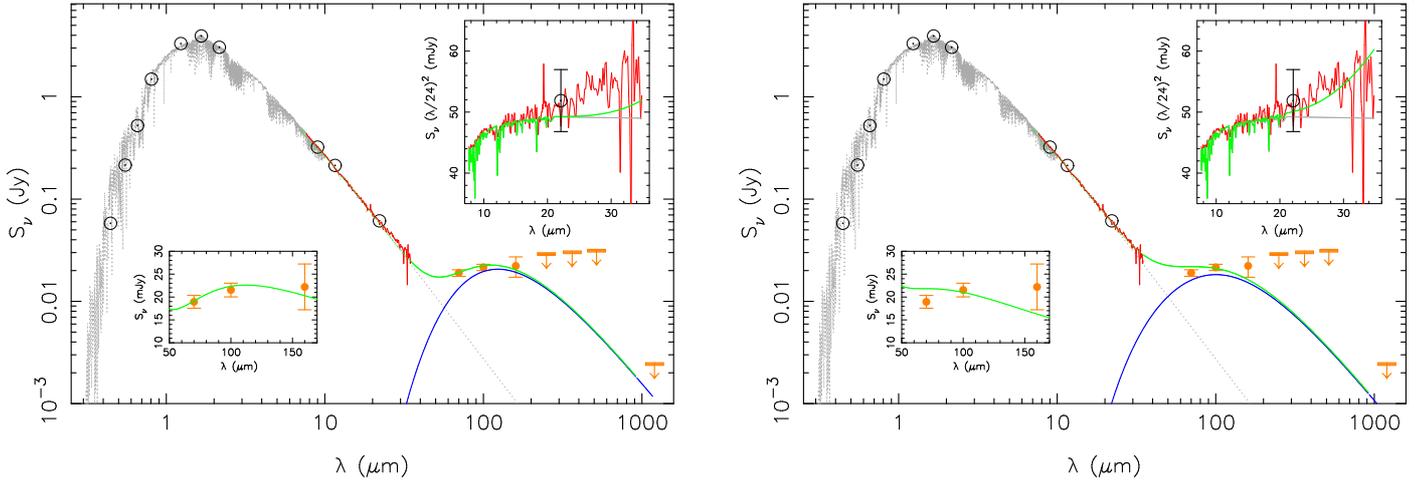

\resizebox{18.5cm}{!}{ \includegraphics[angle=-90]{GJ581_PACS_IRS_26AU.ps} 
\hspace{2.0cm} \includegraphics[angle=-90]{GJ581_PACS_IRS_10AU.ps} }
\caption{SED of the  cold disk model.
The {\it Left-hand figure} shows the best fit to the three PACS images only 
($\chi^2_{\nu}=1.03$, Table~\ref{tab:bestfits}). 
The {\it Right-hand figure} shows the best fit to both the three PACS images and the IRS spectrum  
$((\chi^2_{PACS} + \chi^2_{IRS})/2= 1.05)$.  
The modeled  cold dust emission  is the {\it blue} curve,  
the Next Gen stellar atmosphere spectrum  is the {\it grey} curve, and their sum is  the {\it green} curve.  
The IRS spectrum is in {\it red}.
The upper inset zooms on the IRS wavelengths and displays spectra as $S_{\nu} \times (\lambda/24)^2$  
on a linear scale for clarity. The lower inset zooms on the three PACS bands. In the left-hand 
figure, the best fit model  satisfactorily fits the PACS data as shown by the lower inset but misses 
the IRS data as shown by the upper inset. In the right-hand figure, the best fit model partially misses the
PACS data but  satisfactorily fits the IRS data.}
\label{fig:misfit}
\end{figure*}

We present the SEDs of the star GJ~581 and modeled dust emission that we used to 
determine the fractional dust luminosity $L_{dust}/L_* \sim 10^{-4}$ of the disk. The archival IRS spectrum shows a marginally significant 
excess above the photospheric level that provides additional constraints on the dust emission. However, we show that 
a single cold disk model  and a two component disk model cannot be distinguished to explain this 2-$\sigma$ excess.  

\subsection {The SED and the fractional dust luminosity of the disk} \label{photosphereflx}

The photometry data collected for the SEDs are summarized in Table~\ref{tab:photometry}. The flux densities 
have been color corrected when required 
\footnote{http://herschel.esac.esa.int/twiki/pub/Public/PacsCalibrationWeb/\break cc\_report\_v1.pdf}.
The SED of GJ~581 is based on the NextGen stellar atmospheric model \citep{Haus99}, with the value log(g)=5.0  and  the
effective temperature 3500~K, fit to the Johnson UBV and Cousins RI photometry, 
the $\rm JHK_s$ photometry from 2MASS, and the recent photometry from AKARI and WISE. 
Note that the flux densities of the photosphere used for our modeling in \S~\ref{mod} were predicted from this fit 
(5.8, 2.8 and 1.1~mJy at 70, 100 and 160~$\mu$m, respectively).
In Fig~\ref{fig:misfit} (left-hand panel),  we show this  SED for the star
and the SED for the dust emission from our modeling. 
The  fractional dust luminosity was determined by integrating the SED of 
the dust emission and is $L_{dust}/L_*=8.9 \times 10^{-5}$.  This value is consistent with the fractional dust luminosity 
$Q_{abs}~.~ A~/~4 \pi . r^2 = 9.9 \times 10^{-5}$ determined from the cross-sectional area of the grains $A=2.3~$AU$^2$ 
from our fit in Table~\ref{tab:bestfits}, using the mean disk radius $r=(25+60)/2= 43$~AU, and
assuming the absorption efficiency $Q_{abs}=1$ for the grains larger than 1~$\mu$m. The agreement between
these two independent determinations of the fractional dust luminosity  provides a self-consistency 
check of our modeling. This fractional dust luminosity is higher than that of the Kuiper Belt by several
orders of magnitude.

\subsection{IRS spectrum}

\subsubsection{Synthetic photometry}

The {\it Spitzer} IRS spectrum is  superimposed on the star's SED in Fig~\ref{fig:misfit}. 
As is standard with IRS spectra, the short wavelength module SL ($7.6-14.2~\mu$m) 
has to be adjusted to the predicted photosphere, and  IRS flux densities were scaled up 
by the factor 1.066 for GJ~581.  In Fig~\ref{fig:misfit} and insets, 
 a small excess is apparent above the photospheric level at the longest wavelengths of the spectrum 
(module LL1 : $20.4-34.9~\mu$m). 

 We have carried out synthetic photometry with a rectangular bandpass between 30 and 34~$\mu$m  
and between 15 and 17~$\mu$m  which  gives the widest wavelength range while
still inside of the Long-Low IRS module as  prescribed in \citet{Carp08, Carp09}.
We computed the synthetic flux densities $S_{31.6\mu m}=32.3\pm1.9$~mJy (IRS) and 28.4~mJy (Next~Gen) yielding
the 2$\sigma$ excess $3.9\pm1.9$ mJy, and $S_{15.96\mu m}=110.7\pm0.85$~mJy (IRS) and 109.2~mJy (Next~Gen) 
yielding  the lower significance excess $1.5\pm0.85$~mJy.
We computed  these synthetic flux densities as the weighted mean of the data points in these bands, 
and using the same weights for the corresponding  Next~Gen synthetic flux densities.
The IRS flux density uncertainty includes an absolute calibration error of 6\%. 
Photospheric flux densities predicted for late type stars (K and M) by the Kurucz or Next~Gen models 
 have been shown to be overestimated in the mid-IR  by as much as 3-5\%  
\citep{Gaut07, Lawl09}. Hence, the significance of the marginal
excess at 31.6~$\mu$m  is likely higher in reality.    
If real, this excess for the mature M-star GJ~581 is notable because, even among A-type and solar-type stars,
24~ $\mu$m excesses  are less frequent than  70~$\mu$m excesses 
 and  decrease with age \citep[][]{Riek05, Tril08, Lohn08}. 
In the next two sections, we investigate the implications for the system around GJ~581 
if this excess is real.  

\begin{figure}[!h]
\resizebox{9cm}{!}{\includegraphics[angle=-90]{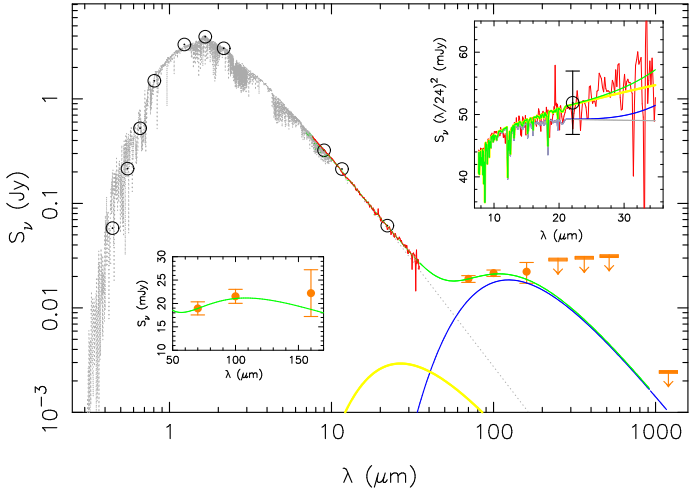}}
\caption{SED of the best fits of  the cold disk and warm belt models. 
 The {\it green} curve corresponds to  the warm and cold dust  
emissions added to the Next Gen stellar
atmosphere spectrum ({\it grey}). IRS spectrum is in {\it red}. 
We also show separately  the  best fit of   
 the  warm belt emission ($r_w=0.2$~AU and $m_w=2.8 \times 10^{-6}$~M$_{\leftmoon}$) ({\it yellow})
and the  best fit  of the cold disk emission (parameters of the combined fit are in Table~\ref{tab:bestfits})  ({\it blue}). 
Insets and photometric data points are the same as in the legend of Fig~\ref{fig:misfit}.}
\label{fig:SED}
\end{figure}

\subsubsection{Modeling the IRS and PACS data with the cold disk model}

First, we fit the single cold disk model of \S~\ref{mod} simultaneously to the three PACS images and the IRS spectrum,
minimizing  $\chi^2_{tot} = (\chi^2_{PACS} + \chi^2_{IRS})/2$ where  $\chi^2_{PACS}$ and $\chi^2_{IRS}$ are the
reduced $\chi^2_{\nu}$ for the  PACS  and IRS data, respectively. With this definition, both data sets 
have the same weight in the fit. The  best fit model thus obtained is characterized by  $\chi^2_{tot}=1.18$, 
resulting from $\chi^2_{IRS}=1.25$  and  $\chi^2_{PACS}=1.11$, and its SED is shown in Fig~\ref{fig:misfit} (right-hand panel). 
The main parameter changes  are $f_T=5.5$ and $A=0.8$~AU$^2$, instead of $f_T=3.5$ and $A=2.3$~AU$^2$ in Table~\ref{tab:bestfits}.   
This value of $\chi^2_{tot}$ is higher than $\chi^2_{\nu}=1.03$  
of the  best fit in this Table and is high for the number of degrees of freedom of 1186 in $\chi^2$-statistics
(probability = 1\% of pure noise).
It is instructive to compare the SEDs of these two fits in Figure~\ref{fig:misfit} ; 
the simultaneous fit to the PACS and IRS data  in the right-hand panel does appear to be skewed to some degree.
The assumption in our current model that the  temperature does not depend on grain size and wavelength
is a limitation. A size distribution would  broaden the  SED,  and it may improve the ability to fit a single disk model to the  flux densities of  the IRS spectrum and the PACS bands simultaneously.

\subsubsection{Modeling the IRS and PACS data with a two component model}

We  explore another possibility, a two component model in which a belt of warm dust 
is added to our cold disk model of \S \ref{mod}. The model of this belt is simply based on  blackbody grains 
(i.e., we set $f_T=1$ for the warm component) located 
at radius $r_w$ and having a total cross-sectional area $A_w$. This two parameter model is fit to  the IRS spectrum alone 
by minimizing  $\chi^2_{IRS}$. We found  $r_w=0.2$~AU ($T_{dust}=191$~K) and    $A_w=7 \times 10^{-5}$~AU$^2$,
giving a corresponding dust mass of $m_w=2.8 \times 10^{-6}$~M$_{\leftmoon}$,  
assuming the standard grain size distribution ($\propto D^{-3.5}$) 
between 1~$\mu$m  and 1~mm-sized particles and $\rho=3$~g/cm$^3$ 
(noting the dependence of this estimate on the unknown 
maximum size given in eq. 4). However, acceptable fits could also be found 
for $r_w$  between 0.05~AU ($T_{dust}=382$~K) and 0.4~AU ($T_{dust}=135$~K), 
encompassing the orbital radii of the planets GJ~581c and GJ~581d.
The IRS data alone cannot constrain  $f_T$  but if this parameter were larger than unity, 
the dust would be $f_T^2$ times further out than the $r_w$ quoted above for the
corresponding $T_{dust}$.     
The SED of  the  two  component model  is shown in Fig~\ref{fig:SED} where we had to decrease 
the cold dust cross-sectional area $A$ by
6\% from its value of Table~\ref{tab:bestfits} to account for the warm dust contributions to the flux densities  
at 70, 100 and 160~$\mu$m. The fractional dust luminosity is 
$L_{dust}/L_*=5.7 \times 10^{-5}$  for  this warm belt shown in yellow in  Fig~\ref{fig:SED}.
Such a belt is  comparable to the  warm disk around the K0 star HD69830 \citep{Liss07}. 
However, the proximity of the warm dust to the known planets suggests that it could 
be dynamically unstable \citep[e.g.][]{Moro07}. Nevertheless, 
definitive proofs are still missing to  establish the  reality of this warm belt in the
GJ~581 system.


\begin{table}[!h]
\caption[]{Photometry of GJ~581}
\label{tab:photometry}
\begin{tabular}{r r l }
\hline\hline
\noalign{\smallskip}
 Wavelength       &  $S_{\nu}$~~~ &  References           \\
  ($\mu$m)~~~~~~~ &  (mJy)        &                   \\
            \noalign{\smallskip}
            \hline
            \noalign{\smallskip}
    0.36      &   $8.3  \pm 2$       &   Hipparcos  \citep{Koen10}        \\
    0.44      &   $61 \pm 10$        &               ~~~~~~~~~~ $('')$    \\     
    0.55      &   $222.7  \pm 3$     &               ~~~~~~~~~~ $('')$    \\
    0.66      &   $523.1  \pm 13$    &               ~~~~~~~~~~ $('')$    \\
    0.81      &   $1490  \pm 14$     &               ~~~~~~~~~~ $('')$    \\
    1.23      & $3317  \pm 82$       &   2MASS   \citep{Cutr03}           \\
    1.66      & $3939  \pm  120$     &               ~~~~~~~~~~ $('')$    \\
    2.16      & $3051  \pm 65$       &               ~~~~~~~~~~ $('')$    \\
    9.0       & $322  \pm 18$        &   AKARI    \citep{Ishi10}          \\
   11.6       & $213   \pm  19$      &   WISE    \citep{Wrig10}           \\
   22.1       & $ 61.2  \pm 6$       &              ~~~~~~~~~~  $('')$    \\
   70.0       & $ 18.9  \pm  1.4$    &   PACS this work                   \\ 
   71.42      & $ 20.0  \pm  5.3$    &   MIPS (*)                         \\ 
  100.0       & $ 21.5  \pm  1.5$    &   PACS this work                  \\
  160.0       & $ 22.2  \pm 5.0$     &              ~~~~~~~~~~  $('')$    \\
  250.0       & $< 24~ ^a$            &    SPIRE this work                 \\
  350.0       & $< 26$  $(3\sigma)$  &              ~~~~~~~~~~  $('')$    \\
  500.0       & $< 27$  $(3\sigma)$  &              ~~~~~~~~~~  $('')$    \\
 1200.0       & $< 2.1$  $(3\sigma)$ &   MAMBO \citep{Lest09}             \\
\noalign{\smallskip}
\hline
\end{tabular}
\null
{\small 
Color correction factors : 1.125 for AKARI, 0.956 for WISE at 11.6~$\mu$m, 0.987 for WISE  at 22.1~$\mu$m, 
and 0.992, 0.980, 0.995 for  PACS (T=40~K) at 70.0, 100.0, 160.0~$\mu$m, respectively. \filbreak
(*) : not used in the fit. \filbreak
$^a$ : see \S~\ref{SPIRE_imag}.}
\end{table}

\section{Brightness limit on scattered light around GJ~581} \label{Nicmos}

\begin{figure}[t!]
\centering
\resizebox{7cm}{!}{\includegraphics[angle=0]{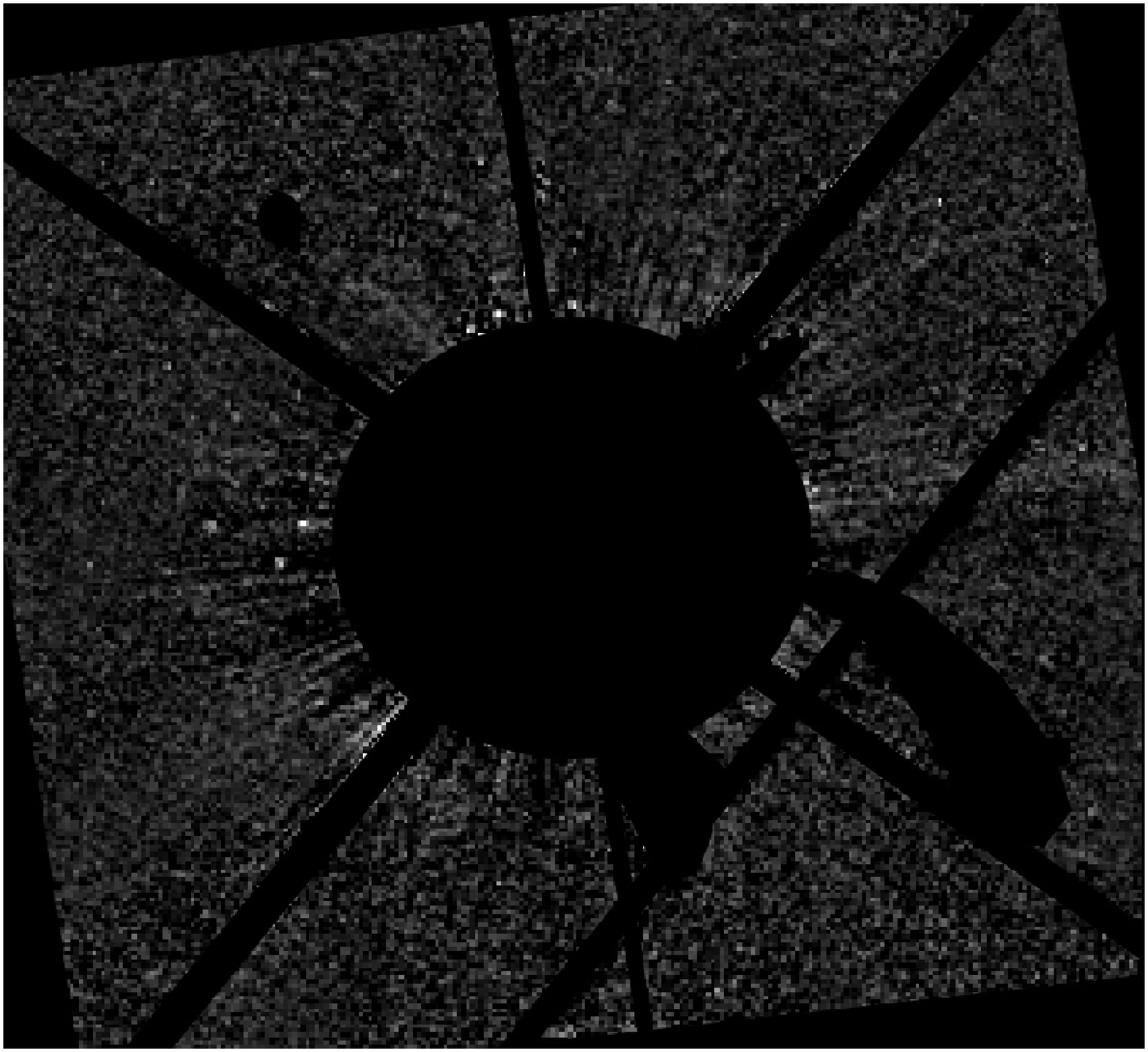}}
\caption{HST/NICMOS F110W image of GJ~581 (North is up, East is left). The star was not
placed behind the occulting spot available on the NIC2 camera.
We subtracted the PSF using observations of GJ~250B
made earlier in the same scientific program, as described in \citet{krist98}.  
The circular black digital mask has 4$''$ radius (30 AU) and blocks the central
region where PSF subtraction artifacts are significant.  Along the
position angle of the disk (PA=120 degrees) the field of view
is limited to approximately 10$''$ radius (62 AU).
We estimate the 3$\sigma$ sensitivity to
nebulosity in the  4$''$-10$''$ radius
region as $\Sigma_{F110W}$~=~18.7~mag~arcsec$^{-2}$.
Lack of detectable scattered light at this level is consistent with the dust model
derived from the far-IR PACS images.}
\label{fig:nicmos}
\end{figure}

Our HST/NICMOS F110W image of GJ~581 after PSF subtraction 
(Fig. \ref{fig:nicmos}) is sensitive to a region 
from 4$''$ radius (30 AU) to approximately 10$''$ radius (62 AU) along PA=120 degrees.
We estimate the 3$\sigma$ sensitivity to nebulosity in this region
as $\Sigma_{F110W}$~=~18.7~mag~arcsec$^{-2}$.
We used the radiative transfer code MCFOST \citep{Pint06} to produce 
synthetic F110W debris disk images for pure astronomical silicate
and pure water ice  models that match the SED based
on the geometry derived in Table~\ref{tab:bestfits}, using
the standard F110W NICMOS throughput. The maximum surface brightnesses
in the $4-10''$ range at the forward scattering peak for pure water ice grains 
and for pure astronomical silicates are  $\sim 19.4$~mag~arcsec$^{-2}$ 
and $\sim 21.2$~mag~arcsec$^{-2}$ , respectively.  Hence, our dust models are
consistent with the non-detection of scattered light around GJ~581, 
but show that the disk may be detectable in deeper observations.

\section{Discussion}  \label{discussion}

We have spatially resolved a disk around the mature M-star GJ~581 hosting four planets. This cold disk is 
reminiscent of the Kuiper Belt in the Solar system  but it surrounds a low mass star ($0.3~M_{\odot}$) 
and has a much higher fractional dust luminosity  $L_{dust}/L_*$ of  $\sim 10^{-4}$.
It shows that  debris disks can survive around M-stars beyond the first tens  of Myr after 
the protoplanetary disk disperses, and they can be detectable although they have been elusive in 
searches so far.

\subsection {Dust temperature in the cold disk}

The factor $f_T$ is significantly larger than unity in our analysis and indicates 
that the dust temperature ranges from $\sim 50$ to $\sim 30$~K from the inner 
to the outer radius  of our modeled  disk. This is about three times the black body 
equilibrium temperature for the dust around this low luminosity M3-type star. 
Values of $f_T$ larger than unity  have also been found for the  debris disks around 
the G-type star 61~Vir  \citep{Wyat12} and several A-type stars \citep{Boot12}. 
This is akin to disks resolved in scattered light which tend to be more extended than
their sizes estimated from blackbody SED \citep{Rodr12}. 
Values $f_T > 1$  are interpreted as evidence for dust grains of small sizes and/or optical properties 
different from blackbody spheres \citep{Back93, Liss07, Bons10}. When the SED of a debris disk 
is finely sampled spectroscopically from mid-IR to submm, a parameter search for composition, 
structure, size distribution of the grains can be conducted usefully \citep[e.g.][]{Lebr12}.
Such a parameter search would be degenerate for GJ~581  because of its limited photometry, and   
so these effects have been reduced to the single parameter $f_T$ in our model.

\subsection {Collision, Poynting-Robertson and stellar wind time
scales for the GJ~581 system}
\label{timescales}

In addition to gravitational forces, dust dynamics
is controlled by radial forces (radiation and stellar wind pressures) and
by tangential forces (Poynting-Robertson and stellar wind drags).
For large dust grains, these perturbing forces act on much longer timescales
than collisions, and such grains simply orbit the star until they are broken
into smaller fragments in collisions with other grains.
This results in a collisional cascade with a size distribution with a
characteristic slope $n(D) \propto D^{-7/2}$ (assuming dust grain strength is
independent of size).
For small dust grains, perturbing forces truncate (or at least significantly
deplete) the size distribution at scales where one of the perturbing
force timescales is shorter than the collisional lifetime \citep[e.g.][]{Wyat11}.
To ascertain the process dominating the dust removal requires a comparison
of the relevant timescales.

\begin{figure}[!t]
\resizebox{9.5cm}{!}{\includegraphics[angle=0]{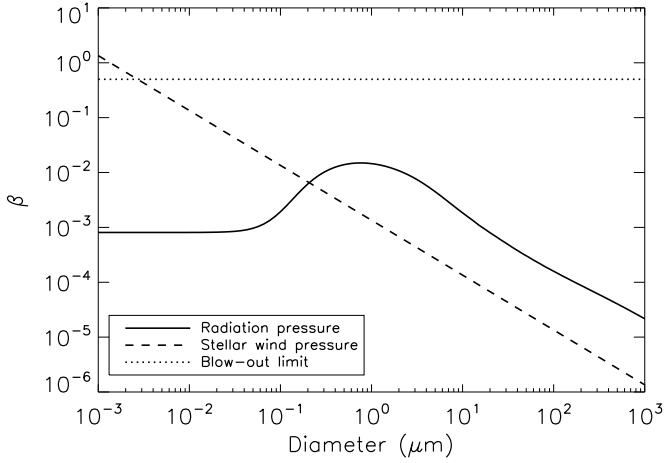}}
\caption{
The ratios of radiation pressure (solid line) and stellar wind pressure
(dashed line) to stellar gravity as a function of particle diameter for
icy grains around GJ~581.
Particles with $\beta>0.5$ are put on hyperbolic orbits as soon as they
are created and so removed from the system on orbital timescales,
thus setting the blow-out limit.
}
\label{fig:beta}
\end{figure}

Fig.~\ref{fig:beta} shows the ratio $\beta$ of the radiation pressure to
stellar gravity experienced by icy dust grains of different sizes.
This peaks for sizes comparable to the wavelength where the stellar
spectrum peaks and is also proportional to $L_*/M_*$, where $L_*$ and $M_*$
are the luminosity and mass of the star \citep{Gust94}.
The low luminosity of GJ~581 means that $\beta < 0.5$ for all icy grains
regardless of size, and the same is true for other compositions.
Since dust with $\beta<0.5$ that is created in collisions is always placed on
a bound orbit, this means that radiation pressure is not a mechanism that can
be invoked to expel the dust from the system \citep{Wyat99}.

Fig.~\ref{fig:beta} also shows the ratio $\beta_{sw}$ of the stellar wind
pressure to stellar gravity \citep{Gust94}.
This depends on the stellar mass loss rate and stellar
wind speed that are poorly understood and hard to measure for M-stars. 
Here we estimate the mass loss rate from the non-detection of X-rays from
GJ~581 by ROSAT implying $\log{L_x} < 26.44$ erg/s \citep{Schm95}.
The correlation between X-ray surface flux and mass loss rate of GKM-type
stars \citep{Wood05} then yields the upper limit of 2~$\dot M_{\odot}$,
where the solar mass loss rate
$\dot M_{\odot} = 2 \times 10^{-14}$~M$_{\odot}$~yr$^{-1}$.
We also consider the stellar wind speed to be $\sim$400 km/s
as appropriate for GKM-type stars \citep{Wood04}.
Finally we assumed 100\% efficiency of momentum coupling between 
dust and the stellar wind (and used eq. 12 of \citet{Plav05}). 
With these assumptions we found that $\beta_{sw}$ can only be $>0.5$ for
dust smaller than a few nm, meaning that stellar wind pressure could only
truncate the collisional cascade below the nm-scale;
furthermore, stellar wind pressure would be ineffective if small grains
couple inefficiently to the stellar wind \citep[e.g.][]{Mina06}. 

A comparison of timescales first requires an estimation of the collisional
lifetime of dust grains of different sizes.
Here we use  eq.~4 and the parameters for the disk found from the modeling of the
combined images presented in Table~\ref{tab:bestfits}  to derive the total mass 
$M_{\rm{tot}}=2.2 \times 10^{-3} \sqrt{{D_c}/10~{\rm cm}}$ in $M_\oplus$,
assuming the standard size distribution between $D_{min}=$1~$\mu$m and 
the diameter $D_c$ of the largest objects and the density $\rho=1.2$~g/cm$^{3}$ for icy grains.
The collisional lifetime is estimated using eq.~16 of \citet{Wyat08} with the
additional assumptions that dust orbital eccentricities are 0.05 and that
their strength is $10^3$ J/kg (independent of size so as to be consistent
with the assumptions about the size distribution).
The resulting collisional lifetime is $0.22\sqrt{D}$ Myr, where $D$ is in $\mu$m.
Since we expect the cascade to extend up to sizes for which their collisional
lifetime is equal to the age of the star ($\sim 5000$ Myr), then 
as long as our assumptions apply up to large sizes we can get a rough estimate
of the total mass of the disk as $0.16M_\oplus$ in objects up to $D_c=0.5$~km
in diameter.

\begin{figure}[!t]
\resizebox{9.5cm}{!}{\includegraphics[angle=0]{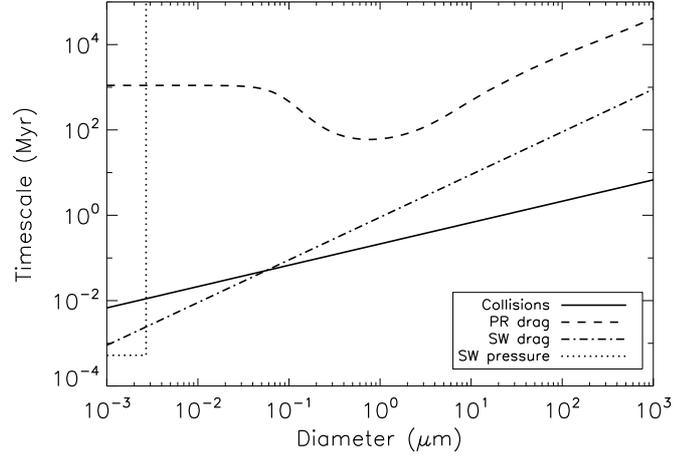}}
\caption{Dust removal timescales as a function of particle size, due to
collisions (solid line), Poynting-Robertson drag (dashed line), 
stellar wind drag (dash-dot line), and stellar wind pressure
(dotted line).}
\label{fig:tpr}
\end{figure}

Fig.~\ref{fig:tpr} shows the timescale for dust to migrate from the inner
edge of the disk at 25~AU to the star due to P-R drag.
The dependence of this timescale on particle size results from a 
scaling $\propto 1/\beta$ which means that this has a minimum value of 60~Myr.
Since this timescale is longer than the collisional lifetime at all sizes,
P-R drag is not a significant loss process from the disc.
Fig.~\ref{fig:tpr} also shows the corresponding timescale for migration
due to stellar wind drag.
This also includes a scaling $\propto 1/\beta_{sw}$, and the efficient
momentum coupling assumed here means that this timescale decreases indefinitely
to smaller sizes $\propto D$.
As a result, stellar wind drag timescales become shorter than collisional timescales
at a size of around 3~nm.

Thus, if all of the assumptions hold, we would expect the collisional cascade
to extend down to 3~nm, while smaller dust is removed by stellar
wind drag.
However, it should be noted that there are significant uncertainties, both
on the magnitude of stellar wind drag and its efficiency of coupling to small
grains, and on the geometry of the dust disk which impacts the collisional
lifetimes.
As such this plot should be considered as representative of the kind of
arguments that need to be considered when assessing the fate of material
in the debris disk of GJ~581.
Further study of this issue is left to later papers, but here we note
that the existence (or not) of grains smaller than 1~$\mu$m is not
important for the observable properties of the disk discussed in this
paper, since such grains are inefficient emitters in the far-IR.

Another scenario that we have not considered in detail is that
the dust is all in large mono-sized grains, in a configuration meaning
that the dust collides at low enough 
velocities that particles bounce off each other rather than destroy each 
other \citep{Heng10}.
Two constraints on such models are that the SED should look like a black
body (since all the dust is large), and the fractional luminosity should
not be large enough that collisions must necessarily occur at high
velocity.
Here the fractional luminosity only constrains the collision velocity
at the inner edge to be $>0.3$m/s which is not sufficient to require
a collisional cascade.
However, although there is no evidence from our limited photometry
of GJ~581 that the spectrum departs from black body shape, the resolved
location of the dust shows that it is significantly hotter than
black body, consistent with the presence of small grains and so incompatible
with this model.

\subsection {Planets and disk relationship for the GJ~581 system}  \label{planets}

First, we note that our determination 
of the inclination of the disk relative to the plane of the sky is  $30^{\circ} < i < 70^{\circ}$
(face on disk is $i=0^{\circ}$). This is mostly constrained by the 160~$\mu$m image and is 
fairly insensitive to the masks used for the background sources.  If the disk mid-plane and the orbits of the planets
are coplanar, this range of inclination
makes the masses of the planets of GJ~581 no more than $\sim 1.6$ times their measured minimum masses by radial velocity
and, interestingly, ensures the long-term stability of the orbits in this system  as shown in dynamical studies 
by \citet{Beus08} and   \citet{Mayo09}.

In our DEBRIS sample of 89 M-stars, there are only three M-stars with known planets  overall
(GJ~876, GJ~832 and GJ~581). 
GJ~581  hosts low mass planets and now has a detected disk, while  GJ~876 and GJ~832  host Jupiter mass planets
and have no detected disk brighter than the fractional dust luminosity $10^{-5}$ in our survey as we shall
present in a future study (Matthews et al. in prep).
Hence, using these three stars as a sample,  the outcome is one disk for 
one low mass planet system (1/1) and no disk for two high mass planet systems (0/2).    
Although this is small number statistics, we note that it is 
suggestive that the correlation between low-mass planets and  debris disks 
recently found  for G-stars by  \citet{Wyat12} also applies to M-stars.
It is also intriguing that the only debris disk confidently detected  in our current analysis  surrounds 
the one star in the sample that hosts low-mass planets.
We note that  simulations by \citet{Raym11, Raym12} suggest that a correlation might exist between low mass 
planets and debris as a result of planet formation processes.
Note that the star AU Mic does not fall in the DEBRIS sample of the nearest M-stars,
and so is not included in the statistics above; this young star (12~Myr) has
a bright disk, but no known planets, although radial velocity measurements 
toward AU Mic are insensitive to planets with masses
lower than a few Jupiters even for short orbital periods because of its high chromospheric activity  
 \citep[see GJ~803 in Fig 19 of][]{Bonf11}. 

Current programs show that a large fraction of M-stars are orbited by low-mass planets.
The radial velocity survey of 102 M-stars  conducted by  \citet{Bonf11} yields the high occurence of 
$35^{+45}_{-11}$\% for low mass planets (2-10~M$_{\oplus}$) around M-stars, 
unlike the low occurence of giant planets of $\sim 2$\%, 
for orbital periods under 100 days. Transit observations in the {\it Kepler} field show that 
small candidate planets ($2-4 R_{\oplus}$) with $P < 50$~days are found around $25 \pm 10$~\% of the M-stars 
($T_{eff}=3600-4100$~K),
seven times more frequently than around the hottest stars (6600-7100 K). There is no such a dependence  
for larger planets  ($4-32 R_{\oplus}$) with $P < 50$~days, found uniformly around  $2\pm1$~\%  
of the stars across all spectral types in the {\it Kepler} field \citep{Howa11}.  
Hence, if there is a correlation between the debris disks and low-mass planets of M-stars 
at a level similar to that found for 
G stars (4/6 nearby G-stars with detectable low-mass planets have 
detectable disks, \citet{Wyat12}, the high fraction of M-stars with low-mass planets
would explain the detection of the disk around GJ 581 but would imply also that 
more disks were expected to be detected in our DEBRIS M-star sample.  We defer this discussion to a paper 
that describes those observations in more detail, since not all 
observations are sensitive to disks at the same level. However, GJ~581 is 
at the median distance of our DEBRIS M-star sample that was observed to
uniform depth, so it could simply be the brightest because of some 
intrinsic properties. The explanation could also be related to its 
multiple planetary system.

Secular perturbation theory has been applied to planetesimals in debris disks perturbed by  planets 
in \citet{Wyat99} and    \citet{Must09}. The outermost planet GJ~581d (5.4~M$_\oplus$, a$_{pl}$=0.22~AU, and e$_{pl}$=0.25, the highest  eccentricity in the system) cannot stir the disk at $a=25$ AU 
because the time scale for orbital crossing of  planetesimals is much longer than the stellar age  \citep[eq. 15 in][]{Must09}. 
However, a hypothetical outer planet, for example a Neptune mass planet (17$_\oplus$) at 5~AU with a moderate orbital eccentricity 
of 0.2, can stir the  disk at $a=25$ AU  in much less than the age of the system, and trigger  destructive collisions 
of 0.5~km-sized bodies \citep[eq. 27 in][]{Must09} to feed a collisional cascade. The most recent detection limit on $m~sini$  from radial velocity measurements  of GJ~581 over 3.3 years indicates that such a planet would not have been detected
\citep[][Fig 13]{Bonf11}, and there is a large region of
parameter space of $m~sini$ vs $a$ over which a planet could both stir the disk 
and have eluded detection in radial velocity measurements.

Alternatively, mild collisions between planetesimals in a weakly  
excited disk could eventually form  a Pluto-sized body,  which in turn stirs the disk so that it produces
dust \citep{Keny08}. This self-stirring scenario is plausible 
since the time scale required for the formation of a Pluto-sized body  at 50AU is
comparable to the age of GJ~581, even when the surface density of solids 
is ten times smaller than the minumim-mass solar nebula around an M3-type star  
\citep[eq. 41][with $x_m=0.1$ ]{Keny08}.

\section{Conclusion}   \label{concl}

We have spatially resolved a debris disk around the M-star  GJ~581
with   {\it Herschel} PACS images at 70, 100 and 160~$\mu$m 
and modeled these observations. 
This  is the second spatially resolved debris disk found around an M-star after AU~Mic, but, in contrast,
GJ~581 is much older and is X-ray quiet.   
Our best fit model is  a  disk, extending radially from $25\pm12$~AU  to more than 60~AU.
Such a cold disk is reminiscent of the Kuiper Belt but it surrounds a low mass star ($0.3~M_{\odot}$) 
and its  fractional dust luminosity  $L_{dust}/L_*$ of  $\sim 10^{-4}$ is much higher.
Also, in our best fit model, the dust temperature is found to be significantly higher than the blackbody equilibrium temperature 
indicating that small grains are abundant. Finally, the inclination limits of the disk make the masses of the planets
small enough to ensure the long-term stability of the system according to dynamical simulations 
by \citet{Beus08} and   \citet{Mayo09}. 

 This disk  complements our view of this remarkable system known to host at least four low mass, close-in planets.
These  planets cannot perturb sufficiently the modeled cold disk to trigger destructive collisions between planetesimals 
over the age of the star, but  a hypothetical outer planet, for example a Neptune mass planet 
with an orbital radius of 5 AU and
a moderate eccentricity, could  replenish the system with
dust. Alternatively,  the self-stirring mechanism could operate for this old star causing sufficient dynamical
excitation to produce the observed dust. 

 It is intriguing that, in our current analysis of the DEBRIS 
 sample of 89 M-stars, the only  debris disk confidently detected around a mature M-star 
also happens to be around the only star known to have low mass planets.  This  could mean that the correlation  between 
low-mass planets and  debris disks recently found for G-stars by \citet{Wyat12}  also applies to M-stars.
Then, the high fraction ($\sim 25\%$) of M-stars known to host low mass planets in the radial velocity  
and {\it Kepler} observations should make debris disks relatively common around them. 
If these disks have not been detected yet, it may be because  searches 
have  simply not been deep enough, or because the disk around GJ~581 is the brightest
owing to some intrinsic properties ; for example hosting a multiple planetary system.   

Future studies and complementary observations of GJ~581 at higher angular resolution will
enhance further our knowledge of  this remarkable system around an M-star.

\begin{acknowledgements}
The Herschel spacecraft was designed, built, tested, and launched under a contract to ESA managed by the Herschel/Planck Project team 
by an industrial consortium under the overall responsibility of the prime contractor Thales Alenia Space (Cannes), 
and including Astrium (Friedrichshafen) responsible for the payload module and for system testing at spacecraft level, 
Thales Alenia Space (Turin) responsible for the service module, and Astrium (Toulouse) responsible for the telescope, 
with in excess of a hundred subcontractors. We thank Ben Zuckerman for comments on a
draft of this article. JFL  gratefully acknowledges  the financial support of Centre National d'Etudes Spatiales (CNES).
JH gratefully acknowledges  the financial support of the Australian government through ARC Grant DP0774000.
MB is funded through a Space Science Enhancement Program
grant from the Canadian Space Agency.
\end{acknowledgements}

\bibliographystyle{aa}
\bibliography{DEBRIS_M056_v4}

\begin{thebibliography}{109}
\expandafter\ifx\csname natexlab\endcsname\relax\def\natexlab#1{#1}\fi

\bibitem[{{Adams} {et~al.}(2004){Adams}, {Hollenbach}, {Laughlin}, \&
  {Gorti}}]{Adam04}
{Adams}, F.~C., {Hollenbach}, D., {Laughlin}, G., \& {Gorti}, U. 2004, \apj,
  611, 360

\bibitem[{{Andrews} \& {Williams}(2005)}]{Andr05}
{Andrews}, S.~M. \& {Williams}, J.~P. 2005, \apj, 631, 1134

\bibitem[{{Augereau} \& {Beust}(2006)}]{Auge06}
{Augereau}, J.-C. \& {Beust}, H. 2006, \aap, 455, 987

\bibitem[{{Aumann} {et~al.}(1984){Aumann}, {Beichman}, {Gillett}, {de Jong},
  {Houck}, {Low}, {Neugebauer}, {Walker}, \& {Wesselius}}]{Auma84}
{Aumann}, H.~H., {Beichman}, C.~A., {Gillett}, F.~C., {et~al.} 1984, \apjl,
  278, L23

\bibitem[{{Avenhaus} {et~al.}(2012){Avenhaus}, {Schmid}, \& {Meyer}}]{Aven12}
{Avenhaus}, H., {Schmid}, H.~M., \& {Meyer}, M.~R. 2012, ArXiv e-prints
  1209.0678

\bibitem[{{Backman} \& {Paresce}(1993)}]{Back93}
{Backman}, D.~E. \& {Paresce}, F. 1993, in Protostars and Planets III, ed.
  {E.~H.~Levy \& J.~I.~Lunine}, 1253--1304

\bibitem[{{Beichman} {et~al.}(2006){Beichman}, {Tanner}, {Bryden},
  {Stapelfeldt}, {Werner}, {Rieke}, {Trilling}, {Lawler}, \&
  {Gautier}}]{Beic06}
{Beichman}, C.~A., {Tanner}, A., {Bryden}, G., {et~al.} 2006, \apj, 639, 1166

\bibitem[{{Berta} {et~al.}(2011){Berta}, {Magnelli}, {Nordon}, {Lutz}, {Wuyts},
  {Altieri}, {Andreani}, {Aussel}, {Casta{\~n}eda}, {Cepa}, {Cimatti}, {Daddi},
  {Elbaz}, {F{\"o}rster Schreiber}, {Genzel}, {Le Floc'h}, {Maiolino},
  {P{\'e}rez-Fournon}, {Poglitsch}, {Popesso}, {Pozzi}, {Riguccini},
  {Rodighiero}, {Sanchez-Portal}, {Sturm}, {Tacconi}, \& {Valtchanov}}]{Bert11}
{Berta}, S., {Magnelli}, B., {Nordon}, R., {et~al.} 2011, \aap, 532, A49

\bibitem[{{Beust} {et~al.}(2008){Beust}, {Bonfils}, {Delfosse}, \&
  {Udry}}]{Beus08}
{Beust}, H., {Bonfils}, X., {Delfosse}, X., \& {Udry}, S. 2008, \aap, 479, 277

\bibitem[{{Blain} {et~al.}(2002){Blain}, {Smail}, {Ivison}, {Kneib}, \&
  {Frayer}}]{Blai02}
{Blain}, A.~W., {Smail}, I., {Ivison}, R.~J., {Kneib}, J.-P., \& {Frayer},
  D.~T. 2002, \physrep, 369, 111

\bibitem[{{Bonfils} {et~al.}(2011){Bonfils}, {Delfosse}, {Udry}, {Forveille},
  {Mayor}, {Perrier}, {Bouchy}, {Gillon}, {Lovis}, {Pepe}, {Queloz}, {Santos},
  {S{\'e}gransan}, \& {Bertaux}}]{Bonf11}
{Bonfils}, X., {Delfosse}, X., {Udry}, S., {et~al.} 2011, ArXiv e-prints
  1111.5019

\bibitem[{{Bonfils} {et~al.}(2005){Bonfils}, {Forveille}, {Delfosse}, {Udry},
  {Mayor}, {Perrier}, {Bouchy}, {Pepe}, {Queloz}, \& {Bertaux}}]{Bonf05}
{Bonfils}, X., {Forveille}, T., {Delfosse}, X., {et~al.} 2005, \aap, 443, L15

\bibitem[{{Bonsor} \& {Wyatt}(2010)}]{Bons10}
{Bonsor}, A. \& {Wyatt}, M. 2010, \mnras, 409, 1631

\bibitem[{{Booth} {et~al.}(2012){Booth}, {Kennedy}, {Sibthorpe}, {Matthews},
  {Wyatt}, {Duch{\^e}ne}, {Kavelaars}, {Rodriguez}, {Greaves}, {Koning},
  {Vican}, {Rieke}, {Su}, {Moro-Mart{\'{\i}}n}, \& {Kalas}}]{Boot12}
{Booth}, M., {Kennedy}, G., {Sibthorpe}, B., {et~al.} 2012, \mnras, 77

\bibitem[{{Broekhoven-Fiene} {et~al.}(2012){Broekhoven-Fiene}, {Matthews},
  {Kennedy}, {Wyatt}, {Duch{\^e}ne}, \& {Kavelaars}}]{Broe12}
{Broekhoven-Fiene}, H., {Matthews}, B., {Kennedy}, G., {et~al.} 2012, \apj

\bibitem[{{Bryden} {et~al.}(2009){Bryden}, {Beichman}, {Carpenter}, {Rieke},
  {Stapelfeldt}, {Werner}, {Tanner}, {Lawler}, {Wyatt}, {Trilling}, {Su},
  {Blaylock}, \& {Stansberry}}]{Bryd09}
{Bryden}, G., {Beichman}, C.~A., {Carpenter}, J.~M., {et~al.} 2009, \apj, 705,
  1226

\bibitem[{{Bryden} {et~al.}(2006){Bryden}, {Beichman}, {Trilling}, {Rieke},
  {Holmes}, {Lawler}, {Stapelfeldt}, {Werner}, {Gautier}, {Blaylock}, {Gordon},
  {Stansberry}, \& {Su}}]{Bryd06}
{Bryden}, G., {Beichman}, C.~A., {Trilling}, D.~E., {et~al.} 2006, \apj, 636,
  1098

\bibitem[{{Carpenter} {et~al.}(2009){Carpenter}, {Bouwman}, {Mamajek}, {Meyer},
  {Hillenbrand}, {Backman}, {Henning}, {Hines}, {Hollenbach}, {Kim},
  {Moro-Martin}, {Pascucci}, {Silverstone}, {Stauffer}, \& {Wolf}}]{Carp09}
{Carpenter}, J.~M., {Bouwman}, J., {Mamajek}, E.~E., {et~al.} 2009, \apjs, 181,
  197

\bibitem[{{Carpenter} {et~al.}(2008){Carpenter}, {Bouwman}, {Silverstone},
  {Kim}, {Stauffer}, {Cohen}, {Hines}, {Meyer}, \& {Crockett}}]{Carp08}
{Carpenter}, J.~M., {Bouwman}, J., {Silverstone}, M.~D., {et~al.} 2008, \apjs,
  179, 423

\bibitem[{{Churcher} {et~al.}(2011){Churcher}, {Wyatt}, {Duch{\^e}ne},
  {Sibthorpe}, {Kennedy}, {Matthews}, {Kalas}, {Greaves}, {Su}, \&
  {Rieke}}]{Chur11}
{Churcher}, L.~J., {Wyatt}, M.~C., {Duch{\^e}ne}, G., {et~al.} 2011, \mnras,
  417, 1715

\bibitem[{{Cutri} {et~al.}(2003){Cutri}, {Skrutskie}, {van Dyk}, {Beichman},
  {Carpenter}, {Chester}, {Cambresy}, {Evans}, {Fowler}, {Gizis}, {Howard},
  {Huchra}, {Jarrett}, {Kopan}, {Kirkpatrick}, {Light}, {Marsh}, {McCallon},
  {Schneider}, {Stiening}, {Sykes}, {Weinberg}, {Wheaton}, {Wheelock}, \&
  {Zacarias}}]{Cutr03}
{Cutri}, R.~M., {Skrutskie}, M.~F., {van Dyk}, S., {et~al.} 2003, VizieR Online
  Data Catalog, 2246, 0

\bibitem[{{Delfosse} {et~al.}(1998){Delfosse}, {Forveille}, {Perrier}, \&
  {Mayor}}]{Delf98}
{Delfosse}, X., {Forveille}, T., {Perrier}, C., \& {Mayor}, M. 1998, \aap, 331,
  581

\bibitem[{{Dent} {et~al.}(2000){Dent}, {Walker}, {Holland}, \&
  {Greaves}}]{Dent00}
{Dent}, W.~R.~F., {Walker}, H.~J., {Holland}, W.~S., \& {Greaves}, J.~S. 2000,
  \mnras, 314, 702

\bibitem[{{Dodson-Robinson} {et~al.}(2011){Dodson-Robinson}, {Beichman},
  {Carpenter}, \& {Bryden}}]{Dods11}
{Dodson-Robinson}, S.~E., {Beichman}, C.~A., {Carpenter}, J.~M., \& {Bryden},
  G. 2011, \aj, 141, 11

\bibitem[{{Engle} \& {Guinan}(2011)}]{Engl11}
{Engle}, S.~G. \& {Guinan}, E.~F. 2011, in Astronomical Society of the Pacific
  Conference Series, Vol. 451, Astronomical Society of the Pacific Conference
  Series, ed. S.~{Qain}, K.~{Leung}, L.~{Zhu}, \& S.~{Kwok}, 285

\bibitem[{{Forbrich} {et~al.}(2008){Forbrich}, {Lada}, {Muench}, \&
  {Teixeira}}]{Forb08}
{Forbrich}, J., {Lada}, C.~J., {Muench}, A.~A., \& {Teixeira}, P.~S. 2008,
  \apj, 687, 1107

\bibitem[{{Forveille} {et~al.}(2011){Forveille}, {Bonfils}, {Delfosse},
  {Alonso}, {Udry}, {Bouchy}, {Gillon}, {Lovis}, {Neves}, {Mayor}, {Pepe},
  {Queloz}, {Santos}, {Segransan}, {Almenara}, {Deeg}, \& {Rabus}}]{Forv11}
{Forveille}, T., {Bonfils}, X., {Delfosse}, X., {et~al.} 2011, ArXiv e-prints
  1109.2505

\bibitem[{{Gautier} {et~al.}(2007){Gautier}, {Rieke}, {Stansberry}, {Bryden},
  {Stapelfeldt}, {Werner}, {Beichman}, {Chen}, {Su}, {Trilling}, {Patten}, \&
  {Roellig}}]{Gaut07}
{Gautier}, III, T.~N., {Rieke}, G.~H., {Stansberry}, J., {et~al.} 2007, \apj,
  667, 527

\bibitem[{{Golimowski} {et~al.}(2004){Golimowski}, {Henry}, {Krist},
  {Dieterich}, {Ford}, {Illingworth}, {Ardila}, {Clampin}, {Franz},
  {Wasserman}, {Benedict}, {McArthur}, \& {Nelan}}]{golimowski04}
{Golimowski}, D.~A., {Henry}, T.~J., {Krist}, J.~E., {et~al.} 2004, \aj, 128,
  1733

\bibitem[{{Golimowski} {et~al.}(2011){Golimowski}, {Krist}, {Stapelfeldt},
  {Chen}, {Ardila}, {Bryden}, {Clampin}, {Ford}, {Illingworth}, {Plavchan},
  {Rieke}, \& {Su}}]{Goli11}
{Golimowski}, D.~A., {Krist}, J.~E., {Stapelfeldt}, K.~R., {et~al.} 2011, \aj,
  142, 30

\bibitem[{{Gordon} {et~al.}(2007){Gordon}, {Engelbracht}, {Fadda},
  {Stansberry}, {Wachter}, {Frayer}, {Rieke}, {Noriega-Crespo}, {Latter},
  {Young}, {Neugebauer}, {Balog}, {Beeman}, {Dole}, {Egami}, {Haller}, {Hines},
  {Kelly}, {Marleau}, {Misselt}, {Morrison}, {P{\'e}rez-Gonz{\'a}lez}, {Rho},
  \& {Wheaton}}]{Gord07}
{Gordon}, K.~D., {Engelbracht}, C.~W., {Fadda}, D., {et~al.} 2007, \pasp, 119,
  1019

\bibitem[{{Greaves} {et~al.}(2005){Greaves}, {Holland}, {Wyatt}, {Dent},
  {Robson}, {Coulson}, {Jenness}, {Moriarty-Schieven}, {Davis}, {Butner},
  {Gear}, {Dominik}, \& {Walker}}]{Grea05}
{Greaves}, J.~S., {Holland}, W.~S., {Wyatt}, M.~C., {et~al.} 2005, \apjl, 619,
  L187

\bibitem[{{Greaves} \& {Wyatt}(2010)}]{Grea10}
{Greaves}, J.~S. \& {Wyatt}, M.~C. 2010, \mnras, 404, 1944

\bibitem[{{Griffin} {et~al.}(2010){Griffin}, {Abergel}, {Abreu}, {Ade},
  {Andr{\'e}}, {Augueres}, {Babbedge}, {Bae}, {Baillie}, {Baluteau}, {Barlow},
  {Bendo}, {Benielli}, {Bock}, {Bonhomme}, {Brisbin}, {Brockley-Blatt},
  {Caldwell}, {Cara}, {Castro-Rodriguez}, {Cerulli}, {Chanial}, {Chen},
  {Clark}, {Clements}, {Clerc}, {Coker}, {Communal}, {Conversi}, {Cox},
  {Crumb}, {Cunningham}, {Daly}, {Davis}, {de Antoni}, {Delderfield}, {Devin},
  {di Giorgio}, {Didschuns}, {Dohlen}, {Donati}, {Dowell}, {Dowell}, {Duband},
  {Dumaye}, {Emery}, {Ferlet}, {Ferrand}, {Fontignie}, {Fox}, {Franceschini},
  {Frerking}, {Fulton}, {Garcia}, {Gastaud}, {Gear}, {Glenn}, {Goizel},
  {Griffin}, {Grundy}, {Guest}, {Guillemet}, {Hargrave}, {Harwit}, {Hastings},
  {Hatziminaoglou}, {Herman}, {Hinde}, {Hristov}, {Huang}, {Imhof}, {Isaak},
  {Israelsson}, {Ivison}, {Jennings}, {Kiernan}, {King}, {Lange}, {Latter},
  {Laurent}, {Laurent}, {Leeks}, {Lellouch}, {Levenson}, {Li}, {Li},
  {Lilienthal}, {Lim}, {Liu}, {Lu}, {Madden}, {Mainetti}, {Marliani}, {McKay},
  {Mercier}, {Molinari}, {Morris}, {Moseley}, {Mulder}, {Mur}, {Naylor},
  {Nguyen}, {O'Halloran}, {Oliver}, {Olofsson}, {Olofsson}, {Orfei}, {Page},
  {Pain}, {Panuzzo}, {Papageorgiou}, {Parks}, {Parr-Burman}, {Pearce},
  {Pearson}, {P{\'e}rez-Fournon}, {Pinsard}, {Pisano}, {Podosek}, {Pohlen},
  {Polehampton}, {Pouliquen}, {Rigopoulou}, {Rizzo}, {Roseboom}, {Roussel},
  {Rowan-Robinson}, {Rownd}, {Saraceno}, {Sauvage}, {Savage}, {Savini},
  {Sawyer}, {Scharmberg}, {Schmitt}, {Schneider}, {Schulz}, {Schwartz},
  {Shafer}, {Shupe}, {Sibthorpe}, {Sidher}, {Smith}, {Smith}, {Smith},
  {Spencer}, {Stobie}, {Sudiwala}, {Sukhatme}, {Surace}, {Stevens}, {Swinyard},
  {Trichas}, {Tourette}, {Triou}, {Tseng}, {Tucker}, {Turner}, {Vaccari},
  {Valtchanov}, {Vigroux}, {Virique}, {Voellmer}, {Walker}, {Ward}, {Waskett},
  {Weilert}, {Wesson}, {White}, {Whitehouse}, {Wilson}, {Winter}, {Woodcraft},
  {Wright}, {Xu}, {Zavagno}, {Zemcov}, {Zhang}, \& {Zonca}}]{Grif10}
{Griffin}, M.~J., {Abergel}, A., {Abreu}, A., {et~al.} 2010, \aap, 518, L3

\bibitem[{{Gustafson}(1994)}]{Gust94}
{Gustafson}, B.~A.~S. 1994, Annual Review of Earth and Planetary Sciences, 22,
  553

\bibitem[{{Hauschildt} {et~al.}(1999){Hauschildt}, {Allard}, {Ferguson},
  {Baron}, \& {Alexander}}]{Haus99}
{Hauschildt}, P.~H., {Allard}, F., {Ferguson}, J., {Baron}, E., \& {Alexander},
  D.~R. 1999, \apj, 525, 871

\bibitem[{{Hawley} {et~al.}(2000){Hawley}, {Reid}, \& {Gizis}}]{Hawl00}
{Hawley}, S., {Reid}, I.~N., \& {Gizis}, J. 2000, in Astronomical Society of
  the Pacific Conference Series, Vol. 212, From Giant Planets to Cool Stars,
  ed. {C.~A.~Griffith \& M.~S.~Marley}, 252

\bibitem[{{Heng} \& {Tremaine}(2010)}]{Heng10}
{Heng}, K. \& {Tremaine}, S. 2010, \mnras, 401, 867

\bibitem[{{H{\o}g} {et~al.}(2000){H{\o}g}, {Fabricius}, {Makarov}, {Urban},
  {Corbin}, {Wycoff}, {Bastian}, {Schwekendiek}, \& {Wicenec}}]{Hog00}
{H{\o}g}, E., {Fabricius}, C., {Makarov}, V.~V., {et~al.} 2000, \aap, 355, L27

\bibitem[{{Howard} {et~al.}(2012){Howard}, {Marcy}, {Bryson}, {Jenkins},
  {Rowe}, {Batalha}, {Borucki}, {Koch}, {Dunham}, {Gautier}, {Van Cleve},
  {Cochran}, {Latham}, {Lissauer}, {Torres}, {Brown}, {Gilliland}, {Buchhave},
  {Caldwell}, {Christensen-Dalsgaard}, {Ciardi}, {Fressin}, {Haas}, {Howell},
  {Kjeldsen}, {Seager}, {Rogers}, {Sasselov}, {Steffen}, {Basri},
  {Charbonneau}, {Christiansen}, {Clarke}, {Dupree}, {Fabrycky}, {Fischer},
  {Ford}, {Fortney}, {Tarter}, {Girouard}, {Holman}, {Johnson}, {Klaus},
  {Machalek}, {Moorhead}, {Morehead}, {Ragozzine}, {Tenenbaum}, {Twicken},
  {Quinn}, {Isaacson}, {Shporer}, {Lucas}, {Walkowicz}, {Welsh}, {Boss},
  {Devore}, {Gould}, {Smith}, {Morris}, {Prsa}, {Morton}, {Still}, {Thompson},
  {Mullally}, {Endl}, \& {MacQueen}}]{Howa11}
{Howard}, A.~W., {Marcy}, G.~W., {Bryson}, S.~T., {et~al.} 2012, \apjs, 201, 15

\bibitem[{{Ishihara} {et~al.}(2010){Ishihara}, {Onaka}, {Kataza}, {Salama},
  {Alfageme}, {Cassatella}, {Cox}, {Garcia-Lario}, {Stephenson}, {Cohen},
  {Fujishiro}, {Fujiwara}, {Hasegawa}, {Ita}, {Kim}, {Matsuhara}, {Murakami},
  {Muller}, {Nakagawa}, {Ohyama}, {Oyabu}, {Pyo}, {Sakon}, {Shibai}, {Takita},
  {Tanab}, {Uemizu}, {Ueno}, {Usui}, {Wada}, {Watarai}, {Yamamura}, \&
  {Yamauchi}}]{Ishi10}
{Ishihara}, D., {Onaka}, T., {Kataza}, H., {et~al.} 2010, VizieR Online Data
  Catalog, 2297, 0

\bibitem[{{Jewitt} {et~al.}(2000){Jewitt}, {Trujillo}, \& {Luu}}]{Jewi00}
{Jewitt}, D.~C., {Trujillo}, C.~A., \& {Luu}, J.~X. 2000, \aj, 120, 1140

\bibitem[{{Kalas} {et~al.}(2005){Kalas}, {Graham}, \& {Clampin}}]{Kala05}
{Kalas}, P., {Graham}, J.~R., \& {Clampin}, M. 2005, \nat, 435, 1067

\bibitem[{{Kalas} {et~al.}(2004){Kalas}, {Liu}, \& {Matthews}}]{kala04}
{Kalas}, P., {Liu}, M.~C., \& {Matthews}, B.~C. 2004, Science, 303, 1990

\bibitem[{{Kennedy} {et~al.}(2012{\natexlab{a}}){Kennedy}, {Wyatt},
  {Sibthorpe}, {Duch{\^e}ne}, {Kalas}, {Matthews}, {Greaves}, {Su}, \&
  {Fitzgerald}}]{Kenn12}
{Kennedy}, G.~M., {Wyatt}, M.~C., {Sibthorpe}, B., {et~al.} 2012{\natexlab{a}},
  \mnras, 421, 2264

\bibitem[{{Kennedy} {et~al.}(2012{\natexlab{b}}){Kennedy}, {Wyatt},
  {Sibthorpe}, {Phillips}, {Matthews}, \& {Greaves}}]{Kenn12b}
{Kennedy}, G.~M., {Wyatt}, M.~C., {Sibthorpe}, B., {et~al.} 2012{\natexlab{b}},
  \mnras, 426, 2115

\bibitem[{{Kenyon} \& {Bromley}(2008)}]{Keny08}
{Kenyon}, S.~J. \& {Bromley}, B.~C. 2008, \apjs, 179, 451

\bibitem[{{Kite} {et~al.}(2011){Kite}, {Gaidos}, \& {Manga}}]{kite11}
{Kite}, E.~S., {Gaidos}, E., \& {Manga}, M. 2011, \apj, 743, 41

\bibitem[{{Koen} {et~al.}(2010){Koen}, {Kilkenny}, {van Wyk}, \&
  {Marang}}]{Koen10}
{Koen}, C., {Kilkenny}, D., {van Wyk}, F., \& {Marang}, F. 2010, \mnras, 403,
  1949

\bibitem[{{K{\'o}sp{\'a}l} {et~al.}(2009){K{\'o}sp{\'a}l}, {Ardila},
  {Mo{\'o}r}, \& {{\'A}brah{\'a}m}}]{Kosp09}
{K{\'o}sp{\'a}l}, {\'A}., {Ardila}, D.~R., {Mo{\'o}r}, A., \&
  {{\'A}brah{\'a}m}, P. 2009, \apjl, 700, L73

\bibitem[{{Krist} {et~al.}(2005){Krist}, {Ardila}, {Golimowski}, {Clampin},
  {Ford}, {Illingworth}, {Hartig}, {Bartko}, {Ben{\'{\i}}tez}, {Blakeslee},
  {Bouwens}, {Bradley}, {Broadhurst}, {Brown}, {Burrows}, {Cheng}, {Cross},
  {Demarco}, {Feldman}, {Franx}, {Goto}, {Gronwall}, {Holden}, {Homeier},
  {Infante}, {Kimble}, {Lesser}, {Martel}, {Mei}, {Menanteau}, {Meurer},
  {Miley}, {Motta}, {Postman}, {Rosati}, {Sirianni}, {Sparks}, {Tran},
  {Tsvetanov}, {White}, \& {Zheng}}]{Kris05}
{Krist}, J.~E., {Ardila}, D.~R., {Golimowski}, D.~A., {et~al.} 2005, \aj, 129,
  1008

\bibitem[{{Krist} {et~al.}(1998){Krist}, {Golimowski}, {Schroeder}, \&
  {Henry}}]{krist98}
{Krist}, J.~E., {Golimowski}, D.~A., {Schroeder}, D.~J., \& {Henry}, T.~J.
  1998, \pasp, 110, 1046

\bibitem[{{Krivov}(2010)}]{Kriv10}
{Krivov}, A.~V. 2010, Research in Astronomy and Astrophysics, 10, 383

\bibitem[{{Lagrange} {et~al.}(2000){Lagrange}, {Backman}, \&
  {Artymowicz}}]{Lagr00}
{Lagrange}, A.-M., {Backman}, D.~E., \& {Artymowicz}, P. 2000, Protostars and
  Planets IV, 639

\bibitem[{{Lawler} {et~al.}(2009){Lawler}, {Beichman}, {Bryden}, {Ciardi},
  {Tanner}, {Su}, {Stapelfeldt}, {Lisse}, \& {Harker}}]{Lawl09}
{Lawler}, S.~M., {Beichman}, C.~A., {Bryden}, G., {et~al.} 2009, \apj, 705, 89

\bibitem[{{Lebreton} {et~al.}(2012){Lebreton}, {Augereau}, {Thi}, {Roberge},
  {Donaldson}, {Schneider}, {Maddison}, {M{\'e}nard}, {Riviere-Marichalar},
  {Mathews}, {Kamp}, {Pinte}, {Dent}, {Barrado}, {Duch{\^e}ne}, {Gonzalez},
  {Grady}, {Meeus}, {Pantin}, {Williams}, \& {Woitke}}]{Lebr12}
{Lebreton}, J., {Augereau}, J.-C., {Thi}, W.-F., {et~al.} 2012, \aap, 539, A17

\bibitem[{{Leggett}(1992)}]{Legg92}
{Leggett}, S.~K. 1992, \apjs, 82, 351

\bibitem[{{Lestrade} {et~al.}(2011){Lestrade}, {Morey}, {Lassus}, \&
  {Phou}}]{Lest11}
{Lestrade}, J.-F., {Morey}, E., {Lassus}, A., \& {Phou}, N. 2011, \aap, 532,
  A120

\bibitem[{{Lestrade} {et~al.}(2006){Lestrade}, {Wyatt}, {Bertoldi}, {Dent}, \&
  {Menten}}]{Lest06}
{Lestrade}, J.-F., {Wyatt}, M.~C., {Bertoldi}, F., {Dent}, W.~R.~F., \&
  {Menten}, K.~M. 2006, \aap, 460, 733

\bibitem[{{Lestrade} {et~al.}(2009){Lestrade}, {Wyatt}, {Bertoldi}, {Menten},
  \& {Labaigt}}]{Lest09}
{Lestrade}, J.-F., {Wyatt}, M.~C., {Bertoldi}, F., {Menten}, K.~M., \&
  {Labaigt}, G. 2009, \aap, 506, 1455

\bibitem[{{Lisse} {et~al.}(2007){Lisse}, {Beichman}, {Bryden}, \&
  {Wyatt}}]{Liss07}
{Lisse}, C.~M., {Beichman}, C.~A., {Bryden}, G., \& {Wyatt}, M.~C. 2007, \apj,
  658, 584

\bibitem[{{Liu} {et~al.}(2004){Liu}, {Matthews}, {Williams}, \&
  {Kalas}}]{Liu04}
{Liu}, M.~C., {Matthews}, B.~C., {Williams}, J.~P., \& {Kalas}, P.~G. 2004,
  \apj, 608, 526

\bibitem[{{L{\"o}hne} {et~al.}(2008){L{\"o}hne}, {Krivov}, \&
  {Rodmann}}]{Lohn08}
{L{\"o}hne}, T., {Krivov}, A.~V., \& {Rodmann}, J. 2008, \apj, 673, 1123

\bibitem[{{Lykawka} {et~al.}(2009){Lykawka}, {Horner}, {Jones}, \&
  {Mukai}}]{Lyka09}
{Lykawka}, P.~S., {Horner}, J., {Jones}, B.~W., \& {Mukai}, T. 2009, \mnras,
  398, 1715

\bibitem[{{Matthews} {et~al.}(2010){Matthews}, {Sibthorpe}, {Kennedy},
  {Phillips}, {Churcher}, {Duch{\^e}ne}, {Greaves}, {Lestrade}, {Moro-Martin},
  {Wyatt}, {Bastien}, {Biggs}, {Bouvier}, {Butner}, {Dent}, {di Francesco},
  {Eisl{\"o}ffel}, {Graham}, {Harvey}, {Hauschildt}, {Holland}, {Horner},
  {Ibar}, {Ivison}, {Johnstone}, {Kalas}, {Kavelaars}, {Rodriguez}, {Udry},
  {van der Werf}, {Wilner}, \& {Zuckerman}}]{Matt10}
{Matthews}, B.~C., {Sibthorpe}, B., {Kennedy}, G., {et~al.} 2010, \aap, 518,
  L135

\bibitem[{{Mayor} {et~al.}(2009){Mayor}, {Bonfils}, {Forveille}, {Delfosse},
  {Udry}, {Bertaux}, {Beust}, {Bouchy}, {Lovis}, {Pepe}, {Perrier}, {Queloz},
  \& {Santos}}]{Mayo09}
{Mayor}, M., {Bonfils}, X., {Forveille}, T., {et~al.} 2009, \aap, 507, 487

\bibitem[{{Minato} {et~al.}(2006){Minato}, {K{\"o}hler}, {Kimura}, {Mann}, \&
  {Yamamoto}}]{Mina06}
{Minato}, T., {K{\"o}hler}, M., {Kimura}, H., {Mann}, I., \& {Yamamoto}, T.
  2006, \aap, 452, 701

\bibitem[{{Morbidelli} {et~al.}(2005){Morbidelli}, {Levison}, {Tsiganis}, \&
  {Gomes}}]{Morb05}
{Morbidelli}, A., {Levison}, H.~F., {Tsiganis}, K., \& {Gomes}, R. 2005, \nat,
  435, 462

\bibitem[{{Moro-Mart{\'{\i}}n} {et~al.}(2007){Moro-Mart{\'{\i}}n}, {Malhotra},
  {Carpenter}, {Hillenbrand}, {Wolf}, {Meyer}, {Hollenbach}, {Najita}, \&
  {Henning}}]{Moro07}
{Moro-Mart{\'{\i}}n}, A., {Malhotra}, R., {Carpenter}, J.~M., {et~al.} 2007,
  \apj, 668, 1165

\bibitem[{{Mustill} \& {Wyatt}(2009)}]{Must09}
{Mustill}, A.~J. \& {Wyatt}, M.~C. 2009, \mnras, 399, 1403

\bibitem[{{Ott}(2010)}]{Ott10}
{Ott}, S. 2010, in Astronomical Society of the Pacific Conference Series, Vol.
  434, Astronomical Data Analysis Software and Systems XIX, ed. {Y.~Mizumoto,
  K.-I.~Morita, \& M.~Ohishi}, 139

\bibitem[{{Petit} {et~al.}(2001){Petit}, {Morbidelli}, \& {Chambers}}]{Peti01}
{Petit}, J.-M., {Morbidelli}, A., \& {Chambers}, J. 2001, \icarus, 153, 338

\bibitem[{{Phillips} {et~al.}(2010){Phillips}, {Greaves}, {Dent}, {Matthews},
  {Holland}, {Wyatt}, \& {Sibthorpe}}]{Phil10}
{Phillips}, N.~M., {Greaves}, J.~S., {Dent}, W.~R.~F., {et~al.} 2010, \mnras,
  403, 1089

\bibitem[{{Pilbratt} {et~al.}(2010){Pilbratt}, {Riedinger}, {Passvogel},
  {Crone}, {Doyle}, {Gageur}, {Heras}, {Jewell}, {Metcalfe}, {Ott}, \&
  {Schmidt}}]{Pilb10}
{Pilbratt}, G.~L., {Riedinger}, J.~R., {Passvogel}, T., {et~al.} 2010, \aap,
  518, L1

\bibitem[{{Pinte} {et~al.}(2006){Pinte}, {M{\'e}nard}, {Duch{\^e}ne}, \&
  {Bastien}}]{Pint06}
{Pinte}, C., {M{\'e}nard}, F., {Duch{\^e}ne}, G., \& {Bastien}, P. 2006, \aap,
  459, 797

\bibitem[{{Plavchan} {et~al.}(2005){Plavchan}, {Jura}, \& {Lipscy}}]{Plav05}
{Plavchan}, P., {Jura}, M., \& {Lipscy}, S.~J. 2005, \apj, 631, 1161

\bibitem[{{Plavchan} {et~al.}(2009){Plavchan}, {Werner}, {Chen}, {Stapelfeldt},
  {Su}, {Stauffer}, \& {Song}}]{Plav09}
{Plavchan}, P., {Werner}, M.~W., {Chen}, C.~H., {et~al.} 2009, \apj, 698, 1068

\bibitem[{{Poglitsch} {et~al.}(2010){Poglitsch}, {Waelkens}, {Geis},
  {Feuchtgruber}, {Vandenbussche}, {Rodriguez}, {Krause}, {Renotte}, {van
  Hoof}, {Saraceno}, {Cepa}, {Kerschbaum}, {Agn{\`e}se}, {Ali}, {Altieri},
  {Andreani}, {Augueres}, {Balog}, {Barl}, {Bauer}, {Belbachir}, {Benedettini},
  {Billot}, {Boulade}, {Bischof}, {Blommaert}, {Callut}, {Cara}, {Cerulli},
  {Cesarsky}, {Contursi}, {Creten}, {De Meester}, {Doublier}, {Doumayrou},
  {Duband}, {Exter}, {Genzel}, {Gillis}, {Gr{\"o}zinger}, {Henning},
  {Herreros}, {Huygen}, {Inguscio}, {Jakob}, {Jamar}, {Jean}, {de Jong},
  {Katterloher}, {Kiss}, {Klaas}, {Lemke}, {Lutz}, {Madden}, {Marquet},
  {Martignac}, {Mazy}, {Merken}, {Montfort}, {Morbidelli}, {M{\"u}ller},
  {Nielbock}, {Okumura}, {Orfei}, {Ottensamer}, {Pezzuto}, {Popesso},
  {Putzeys}, {Regibo}, {Reveret}, {Royer}, {Sauvage}, {Schreiber}, {Stegmaier},
  {Schmitt}, {Schubert}, {Sturm}, {Thiel}, {Tofani}, {Vavrek}, {Wetzstein},
  {Wieprecht}, \& {Wiezorrek}}]{Pogl10}
{Poglitsch}, A., {Waelkens}, C., {Geis}, N., {et~al.} 2010, \aap, 518, L2

\bibitem[{{Raymond} {et~al.}(2011){Raymond}, {Armitage}, {Moro-Mart{\'{\i}}n},
  {Booth}, {Wyatt}, {Armstrong}, {Mandell}, {Selsis}, \& {West}}]{Raym11}
{Raymond}, S.~N., {Armitage}, P.~J., {Moro-Mart{\'{\i}}n}, A., {et~al.} 2011,
  \aap, 530, A62

\bibitem[{{Raymond} {et~al.}(2012){Raymond}, {Armitage}, {Moro-Mart{\'{\i}}n},
  {Booth}, {Wyatt}, {Armstrong}, {Mandell}, {Selsis}, \& {West}}]{Raym12}
{Raymond}, S.~N., {Armitage}, P.~J., {Moro-Mart{\'{\i}}n}, A., {et~al.} 2012,
  \aap, 541, A11

\bibitem[{{Reid} {et~al.}(1995){Reid}, {Hawley}, \& {Gizis}}]{Reid95}
{Reid}, I.~N., {Hawley}, S.~L., \& {Gizis}, J.~E. 1995, \aj, 110, 1838

\bibitem[{{Rieke} {et~al.}(2005){Rieke}, {Su}, {Stansberry}, {Trilling},
  {Bryden}, {Muzerolle}, {White}, {Gorlova}, {Young}, {Beichman},
  {Stapelfeldt}, \& {Hines}}]{Riek05}
{Rieke}, G.~H., {Su}, K.~Y.~L., {Stansberry}, J.~A., {et~al.} 2005, \apj, 620,
  1010

\bibitem[{{Rodriguez} \& {Zuckerman}(2012)}]{Rodr12}
{Rodriguez}, D.~R. \& {Zuckerman}, B. 2012, \apj, 745, 147

\bibitem[{{Schmitt} {et~al.}(1995){Schmitt}, {Fleming}, \& {Giampapa}}]{Schm95}
{Schmitt}, J.~H.~M.~M., {Fleming}, T.~A., \& {Giampapa}, M.~S. 1995, \apj, 450,
  392

\bibitem[{{Selsis} {et~al.}(2007){Selsis}, {Kasting}, {Levrard}, {Paillet},
  {Ribas}, \& {Delfosse}}]{Sels07}
{Selsis}, F., {Kasting}, J.~F., {Levrard}, B., {et~al.} 2007, \aap, 476, 1373

\bibitem[{{Sheppard} \& {Trujillo}(2006)}]{Shep06}
{Sheppard}, S.~S. \& {Trujillo}, C.~A. 2006, Science, 313, 511

\bibitem[{{Sibthorpe} {et~al.}(2012){Sibthorpe}, {Ivison}, {Massey},
  {Roseboom}, {van der Werf}, {Matthews}, \& {Greaves}}]{Sibt12}
{Sibthorpe}, B., {Ivison}, R.~J., {Massey}, R.~J., {et~al.} 2012, ArXiv
  e-prints 1211.0007

\bibitem[{{Smith} \& {Terrile}(1984)}]{Smit84}
{Smith}, B.~A. \& {Terrile}, R.~J. 1984, Science, 226, 1421

\bibitem[{{Smith} {et~al.}(2006){Smith}, {Hines}, {Low}, {Gehrz}, {Polomski},
  \& {Woodward}}]{SmithPS06}
{Smith}, P.~S., {Hines}, D.~C., {Low}, F.~J., {et~al.} 2006, \apjl, 644, L125

\bibitem[{{Strubbe} \& {Chiang}(2006)}]{Stru06}
{Strubbe}, L.~E. \& {Chiang}, E.~I. 2006, \apj, 648, 652

\bibitem[{{Su} {et~al.}(2006){Su}, {Rieke}, {Stansberry}, {Bryden},
  {Stapelfeldt}, {Trilling}, {Muzerolle}, {Beichman}, {Moro-Martin}, {Hines},
  \& {Werner}}]{Su06}
{Su}, K.~Y.~L., {Rieke}, G.~H., {Stansberry}, J.~A., {et~al.} 2006, \apj, 653,
  675

\bibitem[{{Tanner} {et~al.}(2010){Tanner}, {Gelino}, \& {Law}}]{Tann10}
{Tanner}, A.~M., {Gelino}, C.~R., \& {Law}, N.~M. 2010, \pasp, 122, 1195

\bibitem[{{Trilling} {et~al.}(2008){Trilling}, {Bryden}, {Beichman}, {Rieke},
  {Su}, {Stansberry}, {Blaylock}, {Stapelfeldt}, {Beeman}, \&
  {Haller}}]{Tril08}
{Trilling}, D.~E., {Bryden}, G., {Beichman}, C.~A., {et~al.} 2008, \apj, 674,
  1086

\bibitem[{{Udry} {et~al.}(2007){Udry}, {Bonfils}, {Delfosse}, {Forveille},
  {Mayor}, {Perrier}, {Bouchy}, {Lovis}, {Pepe}, {Queloz}, \&
  {Bertaux}}]{Udry07}
{Udry}, S., {Bonfils}, X., {Delfosse}, X., {et~al.} 2007, \aap, 469, L43

\bibitem[{{Vogt} {et~al.}(2012){Vogt}, {Butler}, \& {Haghighipour}}]{Vogt12}
{Vogt}, S.~S., {Butler}, R.~P., \& {Haghighipour}, N. 2012, Astronomische
  Nachrichten, 333, 561

\bibitem[{{Vogt} {et~al.}(2010){Vogt}, {Butler}, {Rivera}, {Haghighipour},
  {Henry}, \& {Williamson}}]{Vogt10}
{Vogt}, S.~S., {Butler}, R.~P., {Rivera}, E.~J., {et~al.} 2010, \apj, 723, 954

\bibitem[{{von Braun} {et~al.}(2011){von Braun}, {Boyajian}, {Kane}, {van
  Belle}, {Ciardi}, {L{\'o}pez-Morales}, {McAlister}, {Henry}, {Jao}, {Riedel},
  {Subasavage}, {Schaefer}, {ten Brummelaar}, {Ridgway}, {Sturmann},
  {Sturmann}, {Mazingue}, {Turner}, {Farrington}, {Goldfinger}, \&
  {Boden}}]{vBra11}
{von Braun}, K., {Boyajian}, T.~S., {Kane}, S.~R., {et~al.} 2011, \apjl, 729,
  L26

\bibitem[{{Wargelin} \& {Drake}(2001)}]{Warg01}
{Wargelin}, B.~J. \& {Drake}, J.~J. 2001, \apjl, 546, L57

\bibitem[{{Wilner} {et~al.}(2012){Wilner}, {Andrews}, {MacGregor}, \&
  {Hughes}}]{Wiln12}
{Wilner}, D.~J., {Andrews}, S.~M., {MacGregor}, M.~A., \& {Hughes}, A.~M. 2012,
  \apjl, 749, L27

\bibitem[{{Wood}(2004)}]{Wood04}
{Wood}, B.~E. 2004, Living Reviews in Solar Physics, 1, 2

\bibitem[{{Wood} {et~al.}(2005){Wood}, {M{\"u}ller}, {Zank}, {Linsky}, \&
  {Redfield}}]{Wood05}
{Wood}, B.~E., {M{\"u}ller}, H.-R., {Zank}, G.~P., {Linsky}, J.~L., \&
  {Redfield}, S. 2005, \apjl, 628, L143

\bibitem[{{Wordsworth} {et~al.}(2011){Wordsworth}, {Forget}, {Selsis},
  {Millour}, {Charnay}, \& {Madeleine}}]{Word11}
{Wordsworth}, R.~D., {Forget}, F., {Selsis}, F., {et~al.} 2011, \apjl, 733, L48

\bibitem[{{Wright} {et~al.}(2010){Wright}, {Eisenhardt}, {Mainzer}, {Ressler},
  {Cutri}, {Jarrett}, {Kirkpatrick}, {Padgett}, {McMillan}, {Skrutskie},
  {Stanford}, {Cohen}, {Walker}, {Mather}, {Leisawitz}, {Gautier}, {McLean},
  {Benford}, {Lonsdale}, {Blain}, {Mendez}, {Irace}, {Duval}, {Liu}, {Royer},
  {Heinrichsen}, {Howard}, {Shannon}, {Kendall}, {Walsh}, {Larsen}, {Cardon},
  {Schick}, {Schwalm}, {Abid}, {Fabinsky}, {Naes}, \& {Tsai}}]{Wrig10}
{Wright}, E.~L., {Eisenhardt}, P.~R.~M., {Mainzer}, A.~K., {et~al.} 2010, \aj,
  140, 1868

\bibitem[{{Wyatt}(2008)}]{Wyat08}
{Wyatt}, M.~C. 2008, \araa, 46, 339

\bibitem[{{Wyatt} {et~al.}(2011){Wyatt}, {Clarke}, \& {Booth}}]{Wyat11}
{Wyatt}, M.~C., {Clarke}, C.~J., \& {Booth}, M. 2011, Celestial Mechanics and
  Dynamical Astronomy, 111, 1

\bibitem[{{Wyatt} {et~al.}(1999){Wyatt}, {Dermott}, {Telesco}, {Fisher},
  {Grogan}, {Holmes}, \& {Pi{\~n}a}}]{Wyat99}
{Wyatt}, M.~C., {Dermott}, S.~F., {Telesco}, C.~M., {et~al.} 1999, \apj, 527,
  918

\bibitem[{{Wyatt} {et~al.}(2012){Wyatt}, {Kennedy}, {Sibthorpe},
  {Moro-Mart{\'{\i}}n}, {Lestrade}, {Ivison}, {Matthews}, {Udry}, {Greaves},
  {Kalas}, {Lawler}, {Su}, {Rieke}, {Booth}, {Bryden}, {Horner}, {Kavelaars},
  \& {Wilner}}]{Wyat12}
{Wyatt}, M.~C., {Kennedy}, G., {Sibthorpe}, B., {et~al.} 2012, \mnras, 424,
  1206

\bibitem[{{Zuckerman}(2001)}]{Zuck01}
{Zuckerman}, B. 2001, \araa, 39, 549

\bibitem[{{Zuckerman} \& {Song}(2004)}]{Zuck04}
{Zuckerman}, B. \& {Song}, I. 2004, \araa, 42, 685

\end{thebibliography}

\end{document}